\documentstyle[12pt,epsfig]{article}

\def\beq{\begin{equation}}
\def\eeq#1{\label{#1}\end{equation}}
\def\eeqn{\end{equation}}
\def\beqa{\begin{eqnarray}}
\def\eeqa#1{\label{#1}\end{eqnarray}}
\def\eeqan{\end{eqnarray}}
\def\CR{\nonumber \\ }
\def\leqn#1{(\ref{#1})}

\def\eps{\epsilon}
\def\to{\rightarrow}

\def\stacksymbols #1#2#3#4{\def\theguybelow{#2}
    \def\vp{\lower#3pt}
    \def\sp{\baselineskip0pt\lineskip#4pt}
    \mathrel{\mathpalette\intermediary#1}}

\def\intermediary#1#2{\vp\vbox{\sp
     \everycr={}\tabskip0pt
     \halign{$\mathsurround0pt#1\hfil##\hfil$\crcr#2\crcr
              \theguybelow\crcr}}}

\begin{document}

\begin{titlepage}

\vskip.5cm
\begin{center}
{\huge \bf Comment on Calculation of Positron} \\
\vskip0.4cm
{\huge \bf  Flux from Galactic Dark Matter} 
\vskip.2cm
\end{center}
\vskip1cm

\begin{center}
{\bf Maxim Perelstein and Bibhushan Shakya} \\
\end{center}
\vskip 8pt

\begin{center}
	{\it Newman Laboratory of Elementary Particle Physics\\
	     Cornell University, Ithaca, NY 14853, USA } \\

\vspace*{0.1cm}

{\tt  mp325@cornell.edu, bs475@cornell.edu}
\end{center}

\vglue 0.3truecm

\begin{abstract}
\vskip 3pt \noindent 
Energetic positrons produced in annihilation or decay of dark matter particles in the Milky Way
can serve as an important indirect signature of dark matter. Computing the positron flux expected in
a given dark matter model involves solving transport equations, which account for interaction of positrons with matter and galactic magnetic fields. Existing calculations solve the equations inside the diffusion
zone, where galactic magnetic fields confine positrons, and assume vanishing positron density on the 
boundaries of this zone. However, in many models, a substantial fraction of the dark matter halo lies outside the diffusion zone. Positrons produced there can then enter the diffusion zone and get trapped, potentially reaching the Earth and increasing the expected flux. We calculate this enhancement
for a variety of models. We also evaluate the expected enhancement of the flux of energetic photons 
produced by the inverse Compton scattering of the extra positrons on starlight and cosmic microwave background. We find maximal flux enhancements of order 20\% in both cases.
\end{abstract}

\end{titlepage}

\section{Introduction}

Multiple observations, ranging from rotation curves of galaxies to the structure of anisotropies in the cosmic microwave background, indicate the presence of substantial amount of dark matter in the universe. While the microscopic nature of dark matter has not been probed so far, one popular hypothesis states that dark matter consists of weakly interacting massive particles (WIMPs). The relic density of WIMPs, predicted within the simplest thermal-relic cosmological scenario, turns out to be consistent with the observed dark matter abundance. This coincidence, and the fact that WIMPs are in fact predicted by many popular extensions of the standard model of particle physics, make the WIMP hypothesis quite attractive on theoretical grounds. 

If dark matter in the Milky Way halo indeed consists of WIMPs, occasional pair-annihilation of these particles should produce spectacular high-energy cosmic rays, giving a potential indirect signature for dark matter. In particular, energetic positrons produced in WIMP annihilation events provide a promising way to look for dark matter, which has been exploited by a number of experiments. Two satellite-borne experiments, HEAT~\cite{HEAT} and PAMELA~\cite{Pamela}, provided measurements of the positron flux in the $1-100$ GeV range. In addition, several recent experiments - most notably ATIC~\cite{ATIC}, FERMI~\cite{FERMI}, and HESS~\cite{HESS} - have measured the sum of electron and positron fluxes, extending to TeV energies. All of these experiments report fluxes and spectra inconsistent with ``canonical" background models. The excess fluxes can be interpreted as coming from dark matter annihilation; while the minimal WIMP models tend to predict fluxes too low to be consistent with experiments, simple extensions of the WIMP paradigm, {\it e.g.} incorporating the Sommerfeld mechanism to enhance the annihilation rates at low velocities~\cite{Sommer}, can fit the data. A decaying dark matter particle with lifetime of order $10^{26}$ sec is another candidate~\cite{DecayDM}. It should be noted that many uncertainties remain in the evaluation of positron fluxes from conventional astrophysical sources, and viable explanations of the observed excesses in terms of conventional astrophysical sources have been proposed~\cite{pulsars,Vahe}. Still, the recent rapid experimental progress in this area has highlighted the need for accurate predictions of positron fluxes, both from conventional and unconventional ({\it e.g.} dark matter) sources.

Our focus in this paper is on the calculation of positron fluxes from dark matter annihilation. The calculation proceeds in two steps. First, the spectrum of positrons emerging from a WIMP pair-annihilation event is calculated. (This spectrum depends on the particle physics model responsible for WIMPs: for example, WIMPs can annihilate directly into $e^+e^-$ pairs, or into $W^+W^-$ with subsequent decay $W^+\to e^+\nu_e$, etc.) Second, the interactions of the positrons with galactic magnetic fields, starlight and CMB photons, synchrotron radiation, and other effects occurring on the way from the production point to the detector must be included. The propagation of positrons through the galactic medium is governed by the transport equations, generally a complicated system of coupled differential equations involving densities of positrons, photons and other cosmic ray species. 
Comprehensive, detailed description of galactic propagation requires numerical techniques, and extensive packages such as {\tt GALPROP}~\cite{GALPROP} have been developed to tackle the problem. However, a reasonable first approximation to positron propagation can be obtained by treating the positron density in isolation, and modeling positron interactions with the medium by simple linear diffusion and energy-loss terms (see e.g. Refs.~\cite{Green,Bessel}). This results in the ``diffusion-loss" equation:
\beq
\frac{\partial\psi}{\partial t} \,-\, \nabla\,\left[ K({\bf x}, E)\,\nabla \psi \right]\,-\frac{\partial}{\partial E} \,\left[ b(E) \psi\right] \,=\, q({\bf x}, E)\,,
\eeq{transport}
where $\psi({\bf x}, E, t)=dn_{e^+}/dE$ is the positron density per unit volume per unit energy; $K$ is the diffusion coefficient describing the interaction of the positron with the stochastic galactic magnetic field; $b$ is the rate at which positrons lose energy due to synchrotron emission and inverse Compton scattering (ICS); and $q$ is the source term, in our case due to dark matter annihilation or decay. It is reasonable to assume that the present positron density reflects a time-independent steady state, and solve Eq.~\leqn{transport} with $\partial \psi/\partial t=0$. To do this, boundary conditions in space need to be imposed. 
Galactic magnetic fields are confined to a cylindrical ``diffusion region" or ``diffusion zone", with radius $R$ of order 20 kpc and half-thickness $L$ taken to be between 1 and 15 kpc. If a positron is injected inside the diffusion region, it slowly random-walks through the region, taking of order $10^8$ years (for typical parameters) to reach the edge. Once the edge is reached, the positron escapes into intergalactic medium, traveling essentially with the speed of light. Thus, in steady state, the positron density outside the diffusion region is expected to be strongly suppressed, and flux calculations assume $\psi=0$ at the diffusion region boundaries, $z=\pm L$ and $r=R$. The transport equation is then solved within this cylindrical region. Essentially all existing calculations of positron flux from dark matter annihilation, analytic or numerical,
make this assumption.\footnote{Some of the uncertainties in dark matter indirect signals associated with modeling the diffusion zone boundary have been recently studied in Ref.~\cite{strumia}, using the model with a position-dependent (exponential in $z$) diffusion coefficient proposed in~\cite{Grasso}. Also, the positron flux from galactic {\it subhalos} has been considered in Ref.~\cite{Cline}.}
The key observation of this paper is that this choice of boundary conditions results in an under-estimate of the positron fluxes from dark matter annihilation. 

\begin{figure}[t]
\centering
\begin{tabular}{cc}
\includegraphics[width=3in,keepaspectratio]{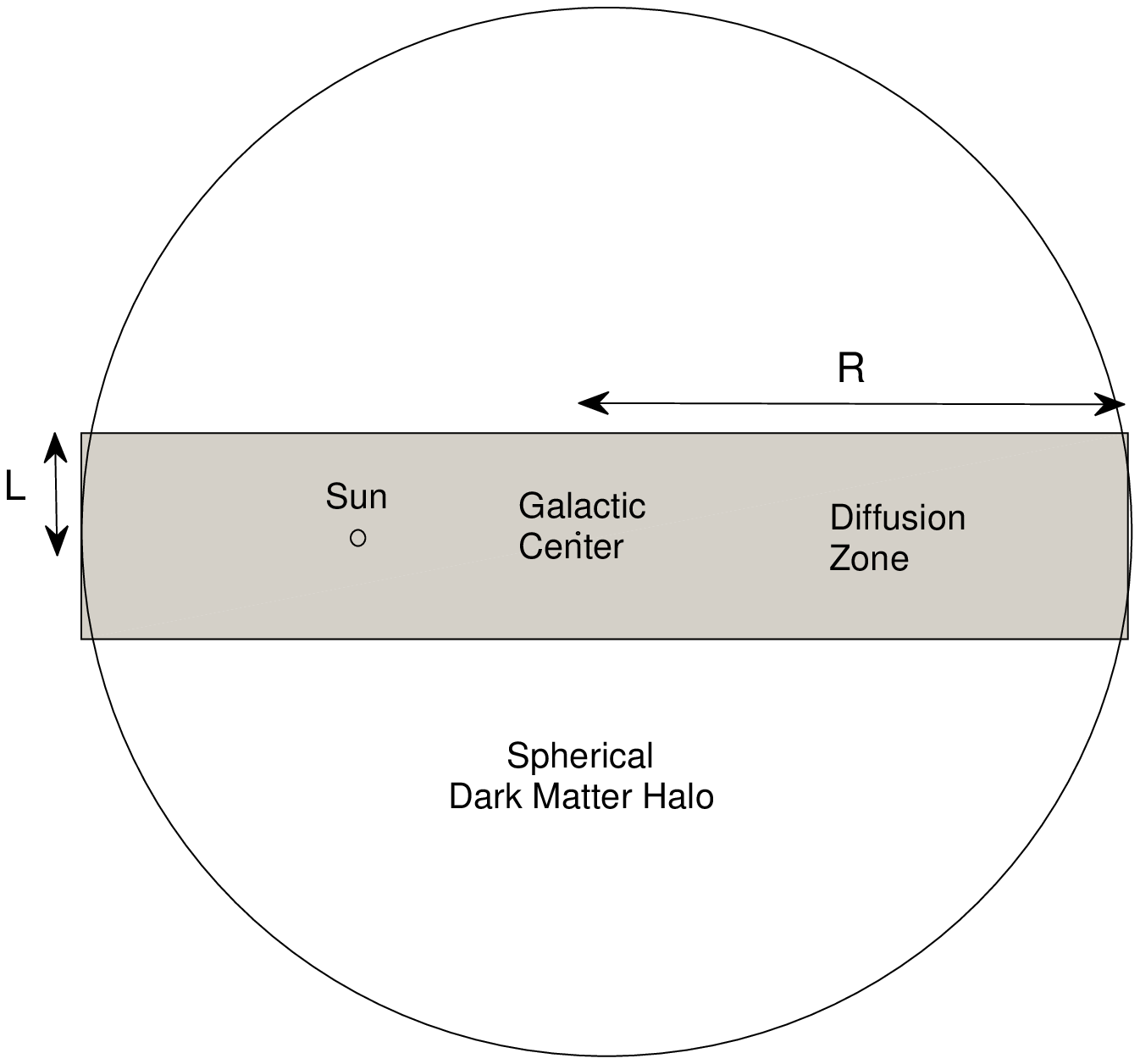}&
\includegraphics[width=3in,keepaspectratio]{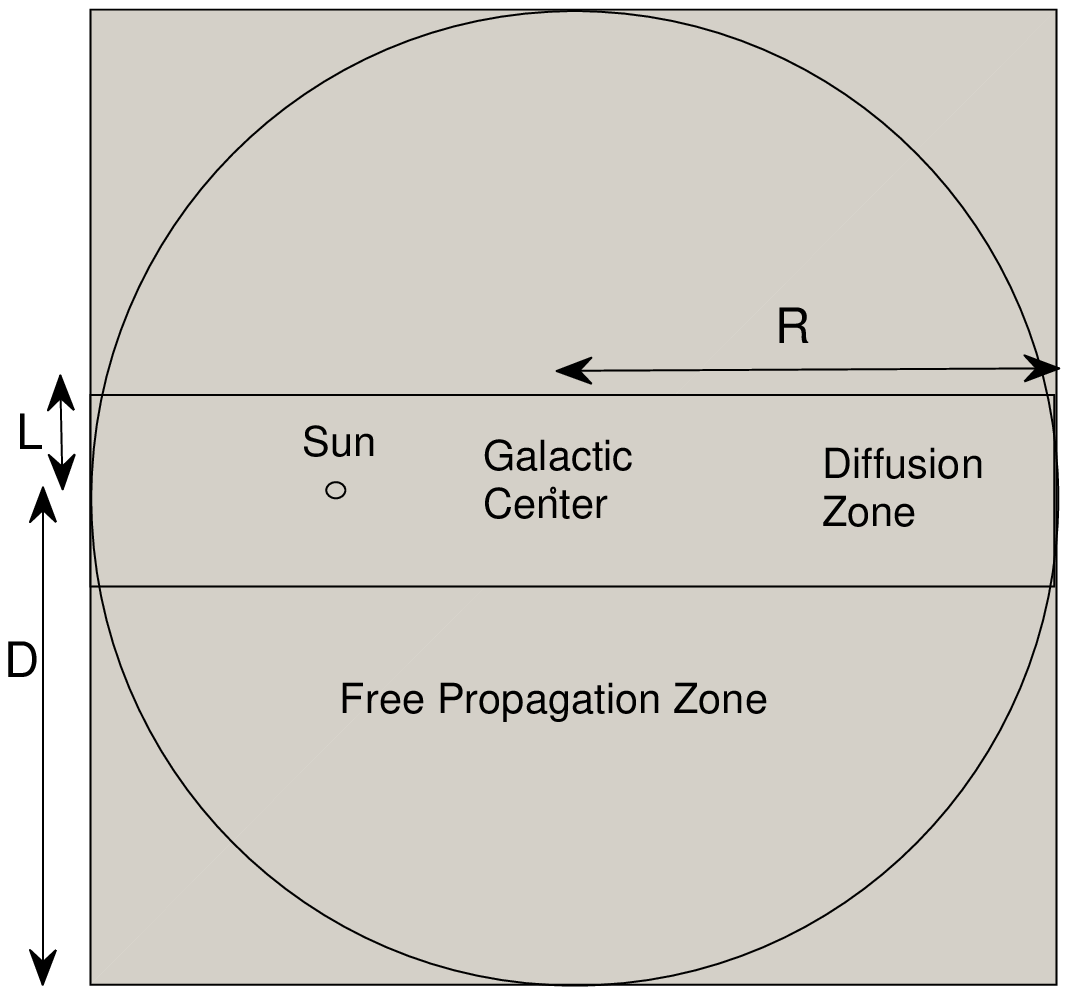}\\
\end{tabular}
\caption{Left -- The dark matter halo extends significantly beyond the diffusion zone, but only sources inside the zone are considered in the conventional formalism (Section \ref{sec:oldform}). Right -- The extended formalism (Section \ref{sec:newform}) considers sources in the free propagation zone in addition to the diffusion zone.}
\label{fig:geometry}
\end{figure}

The source of the flux enhancement is illustrated in Fig.~\ref{fig:geometry}. A typical dark matter halo is spherically symmetric and extends {\it beyond} the diffusion region, in particular in the vertical direction: for example, for an isothermal dark matter profile in the M2 propagation model (which uses $L=1$ kpc), the diffusion zone contains only 10\% of the dark matter mass of the full halo. An order-one fraction of the positrons produced by dark matter annihilations outside the diffusion region will enter the diffusion region, and get ``stuck" there. These positrons will contribute to the steady-state density (and flux) inside the diffusion region. This contribution is missing from any calculation that only considers the sources at $|z|\leq L$. It cannot be incorporated by simply extending $L$ to a larger value since, for a given propagation model, $L$ is a fixed physical parameter that represents how far the galactic magnetic fields extend in space. What is needed is a formalism that keeps galactic magnetic fields confined within the region defined by $L$ as required by the propagation model being used, yet incorporates contributions from dark matter annihilation beyond this region. In this paper we present such a formalism, via a simple extension of the Bessel-transform approach of Ref.~\cite{Bessel}. We then analyze its impact quantitatively, in a number of illustrative models. We also analyze the contribution of the extra positrons to the expected flux of energetic photons from the inverse Compton scattering (ICS) process in the galaxy, suggested in Ref.~\cite{CRspectra} as a robust signature of strongly annihilating dark matter.

\section{Formalism}

In this section, we will first review the conventional Bessel-transform approach~\cite{Bessel} to solving Eq.~\leqn{transport}, and then
present our extension of this formalism.

\subsection{Conventional Formalism}
\label{sec:oldform}

In the conventional formalism, one assumes that the diffusion coefficient and the energy loss term are position-independent within the diffusion zone, but depend on energy. To describe the energy dependence, we adopt the simple power-law parametrization of 
Ref.~\cite{Bessel}:
\beq
K(E) = K_0 \eps^\delta,~~~b(E) = \frac{E_0}{\tau_E}\,\eps^2\,,
\eeq{Edep}
where $\eps=E/E_0$, $E_0=1$ GeV. The quantities $K_0$, $\delta$, and $\tau_E$, along with $R$ and $L$, define the ``galactic propagation model". Standard choices are $\tau_E=10^{16}$ sec and $R=20$ kpc. We will use several combinations of the other parameters, summarized in Table~\ref{tab:galprop}. These combinations have been found compatible with observed cosmic ray properties, such as the B/C ratio~\cite{BC}. 

Assuming that the positron source term is cylindrically symmetric,
and imposing the boundary condition $\psi=0$ at $r=R$ and $z=\pm L$, the solution to Eq.~\leqn{transport} can be expressed as a Bessel-Fourier series:
\beq
\psi(z,r,\epsilon)=\sum_{i}\sum_{n} P_{i,n}(\epsilon)J_0\left(\frac{\alpha_ir}{R}\right)\sin\left(\frac{n\pi(z+L)}{2L}\right)\,,
\eeq{Bessel}
where $J_i$ denotes the $i$-th order Bessel function of the first kind, and $\alpha_{i}$'s are the zeros of $J_0$. Taking the Bessel and Fourier transforms of Eq.~\leqn{transport} at steady state and changing variables according to 
\beq
t = \frac{\tau_E\epsilon^{\delta-1}}{1-\delta}\,,~~~ \tilde{P}_{i,n}=\epsilon^2P_{i,n}
\eeq{change}
gives:
\beq
\frac{d\tilde{P}_{i,n}}{dt} + K_0\left(\left(\frac{\alpha_i}{R}\right)^2+\left(\frac{n\pi}{2L}\right)^2\right)\tilde{P}_{i,n} = \epsilon^{2-\delta}Q_{i,n}\,,
\eeq{ODE}
where the coefficients $Q_{i,n}$ are the Bessel-Fourier transforms of 
the source term: 
\beq
Q_{i,n}(\eps)=\frac{2}{J_1(\alpha_i)^2LR^2}\int^R_0rdr\int^L_{-L}dzJ_0\left(\frac{\alpha_ir}{R}\right)\sin\left(\frac{n\pi(z+L)}{2L}\right)\,q(r,z,\eps)\,.
\eeq{Qs}
This equation can be easily solved:
\beq
\tilde{P}_{i,n}(t)\,=\,\int_0^t\tilde{Q}_{i,n}(t_S)\,\exp\Bigl[-
\omega_{i,n} (t-t_S) \Bigr]\,dt_S\,,
\eeq{ODEsol}
where $\tilde{Q}_{i,n} = \eps^{2-\delta} Q_{i,n} = \left(\frac{\tau_E}{(1-\delta)t}\right)^{\frac{\delta-2}{\delta-1}} Q_{i,n}$, and
\beq
\omega_{i,n} \,=\, K_0\left[ \left(\frac{\alpha_i}{R}\right)^2+\left(\frac{n\pi}{2L}\right)^2 \right]\,. 
\eeq{ODEsol}
In dark matter applications, the positron source energy spectrum is position-independent: $q(r,z,\eps) =R(r,z) f(\eps)$, where $f(\eps)=dN_{e^+}/dE_{e^+}$ is the energy distribution of positrons from a single WIMP annihilation or decay. Specifically, for annihilating dark matter, $R(r,z) =\eta\langle\sigma v\rangle\left(\frac{\rho(r,z)}{m_\chi}\right)^2$, where $\eta$ is equal to 1/2 for Majorana WIMPs and 1/4 for Dirac WIMPs, $\rho$ is dark matter energy density, $m_\chi$ is the WIMP mass, and $\langle \sigma v \rangle$ is the thermally averaged cross section for WIMPs to annihilate into a final state containing positrons. For decaying dark matter, $R(r,z) = \Gamma\rho(r,z)/m_\chi$, where $\Gamma$ is the partial decay width into a final state containing positrons. 
For such sources, the positron density can be expressed as
\beq
\psi(r,z,\eps) \,=\, \frac{\tau_E}{\eps^2} \,\int_\eps^{\eps_{\rm max}}
d\eps_S\,f(\eps_S)\,I(r,z,\eps, \eps_S)\,,
\eeq{Psi_sol}
where $\eps_{\rm max}$ is the maximum energy at which positrons are produced (typically equal to $m_\chi$), and 
\beq
I(r,z,\eps, \eps_S) \,=\,\sum_{i}\sum_{n} J_0\left(\frac{\alpha_ir}{R}\right)\sin\left(\frac{n\pi(z+L)}{2L}\right)\
\exp \Bigl[-\omega_{i,n} (t-t_S) \Bigr] R_{i,n}  
\eeq{hfn}
is the {\it halo function}. Here $R_{i,n}$ are coefficients of the Bessel-Fourier expansion of the function $R(r,z)$. Note that the halo function only depends on $\eps$ and $\eps_S$ through the combination
\beq
t - t_S = \frac{\tau_E}{1-\delta}\left(\eps^{\delta-1}-\eps_S^{\delta-1}\right) \,=\, \frac{\lambda_D^2}{4K_0}\,,
\eeq{dl}
where $\lambda_D$ is the {\it diffusion length}. 

\subsection{Extended Formalism}  
\label{sec:newform}

To include contributions from dark matter annihilations or decays outside the diffusion zone, we would like to solve Eq.~\leqn{transport}
in a larger cylinder extending out to $|z|=D$, with $D>L$ such that all (or essentially all) of the dark matter is contained within this cylinder. We impose the boundary condition $\psi=0$ at $z=\pm D$, since there is no influx of positrons through the boundaries. Inside the cylinder, there are two zones: {\it diffusion zone}, extending out to $|z|=L$, and {\it free-propagation zone}, with $L<|z|<D$. We expect that the positron density is small in the free-propagation zone, since the positrons propagate through this zone very quickly; however, there is an influx of positrons from the free-propagation zone into the diffusion zone, which increases the steady state flux there. For mathematical convenience, we model the propagation through the free-propagation zone by the same diffusion-loss equation, Eq.~\leqn{transport}, but with a different diffusion coefficient $K_e=K_1\eps^\delta$. Ideally we would like to take the limit $K_1\to\infty$, corresponding to very fast diffusion. In practice, numerical convergence issues put an upper limit on the ratio of $K_1/K_0$; however, we are able to calculate with values of $K_1$ for which the results are essentially independent of that parameter, suggesting that the limiting behavior has been reached. In principle, we would also like to set the energy loss term to zero in the free-propagation zone, $b\to 0$. However, since the positrons spend negligible time in this zone, in practice this term does not affect the results, and for simplicity we use the same $b$ in the diffusion and free-propagation zones. 

With a position-dependent diffusion coefficient, the steady state transport equation is 
\beq
-\nabla K\cdot \nabla\psi - K \, \Delta\psi - \frac{\partial}{\partial E}[b({\bf x}, E)\psi] = q({\bf x},E)\,.
\eeq{transport1}
Since both $K$ and $q$ have cylindrical symmetry, we again expand  
$\psi$ and $q$ in Bessel-Fourier series as in Eqs.~\leqn{Bessel} and~\leqn{Qs}, but with $L\to D$. Substituting these expansions into Eq.~\leqn{transport1} yields (for each $i$):
\beqa
& & K(z, \eps) \sum_n P_{i,n} (\eps) \left(\left(\frac{\alpha_i}{R}\right)^2+\left(\frac{n\pi}{2D}\right)^2\right)\sin\left(\frac{n\pi(z+D)}{2D}\right)\CR
&-&\frac{\partial K}{\partial z} \sum_n P_{i,n} (\epsilon) \left(\frac{n\pi}{2D}\right)\cos\left(\frac{n\pi(z+D)}{2D}\right)  \CR
&-& \frac{1}{\tau_E} \frac{\partial}{\partial \eps}\left(\epsilon^2\sum_n P_{i,n}(\eps) \sin\left(\frac{n\pi(z+D)}{2D}\right)\right) = \sum_n Q_{i,n}(\eps) \sin\left(\frac{n\pi(z+D)}{2D}\right)\,.
\eeqa{tr2}
In our setup, the diffusion coefficient has the form  
\beq
K(z,\eps) = \left(K_0 + \tilde{K}(z) \right) \eps^\delta\,,
\eeq{Ktilde}
where $\tilde{K}\to 0$ in the diffusion zone and $\tilde{K}\to K_1-K_0 \approx K_1$ in the free-propagation zone. Multiplying both sides of Eq.~\leqn{tr2} by $\sin\left(\frac{m\pi(z+D)}{2D}\right)$ (where $m$ is an integer) and integrating over $z\in[-D,D]$ gives 
\beqa
&-& \sum_n \tilde{P}_{i,n} \left(\frac{n\pi}{2D^2}\right)\,\int_{-D} ^D \frac{d\tilde{K}}{d z} \cos\left(\frac{n\pi(z+D)}{2D}\right)\sin\left(\frac{m\pi(z+D)}{2D}\right)dz \CR
&+& \frac{1}{D} \sum_n \tilde{P}_{i,n} \left(\left(\frac{\alpha_i}{R}\right)^2+\left(\frac{n\pi}{2D}\right)^2\right)
\int_{-D}^D \tilde{K}(z)\sin\left(\frac{n\pi(z+D)}{2D}\right)\sin\left(\frac{m\pi(z+D)}{2D}\right)dz \CR
&+& K_0\left(\left(\frac{\alpha_i}{R}\right)^2+\left(\frac{m\pi}{2D}\right)^2\right)\tilde{P}_{i,m}+  \, \frac{d}{dt} \tilde{P}_{i,m} = \tilde{Q}_{i,m}\,,
\eeqa{tr3}
where we performed a change of variables as in Eq.~\leqn{change}.
The last line is just what one would obtain in the conventional formalism with $L\to D$; however, there are now additional terms that mix different Fourier components of the positron density. It is useful to put this system of equations in matrix form:
\beq
\frac{d\textbf{P}_i}{dt}+\textbf{A}_i\cdot\textbf{P}_i = \textbf{Q}_i\,,
\eeq{tr_mat}
where \textbf{P}$_{i}$ and \textbf{Q}$_{i}$ are vectors containing the $\tilde{P}_{i,n}$ and $\tilde{Q}_{i,n}$, and  $\textbf{A}_i$ are matrices whose elements can be read from Eq.~\leqn{tr3}. The matrices $\textbf{A}_i$ are $t$-independent.  The solution is given by
\beq
\textbf{P}_i(t) = \int_0^t dt_S \,\exp\Bigl[ -(t-t_s) {\textbf A}_i \Bigr]\, {\textbf Q}_i \,.
\eeq{tr_solve}
The positron density from dark matter decay or annihilation has the same form as in the conventional formalism, Eq.~\leqn{Psi_sol}, with the halo function~\leqn{hfn} replaced by
\beq
I(r,z,\eps, \eps_S) \,=\,\sum_{i}\sum_{n} J_0\left(\frac{\alpha_ir}{R}\right)\sin\left(\frac{n\pi(z+D)}{2D}\right)\
\left( \exp \Bigl[-(t-t_S) {\textbf A}_i \Bigr] {\textbf R}_{i} \right)_n  \,.
\eeq{hfn_new} 
As before, the halo function only depends on $\eps$ and $\eps_S$ through $t-t_S$; however, in this case, the definition of diffusion length $\lambda_D$ is ambiguous, since it involves the diffusion coefficient which is now $z$-dependent. Below, we will always use the value of the diffusion coefficient {\it inside} the diffusion zone to define $\lambda_D$, in complete analogy with Eq.~\leqn{dl}. 

To proceed with the analysis, we need to specify precisely how the diffusion coefficient depends on $z$. The simplest choice is to model it as a step function with the conventional value $K_0$ in the diffusion zone and a much larger value $K_1$ in the free-propagation zone. Such a model, however, is ill-suited for the Bessel-Fourier approach, since an infinitely sharp jump in $K$ requires a very large number of terms in the expansion to achieve convergence. To avoid this problem, we smooth out the step function across a finite interval $L\leq z \leq L+d$, with $d\ll L, D$. Specifically, we assume
\beq
K(z) \,=\, \cases{K_0\,,&if $ |z|\leq L$;\cr\cr
\frac{1}{2}(K_1+K_0)-\frac{1}{4}(K_1-K_0)\Bigl[3\cos\left(\frac{|z|-L}{d}\pi\right)-\cos^3\left(\frac{|z|-L}{d}\pi\right)\Bigr]\,,&if $L<|z|\leq L+d$;\cr\cr
K_1\,,&if $L+d<|z|\leq D$\,.}
\eeq{Kz} 
This function has continuous first and second derivatives at $z=L$ and $z=L+d$. Another important advantage of this form is that all integrals in Eq.~\leqn{tr3} can be evaluated analytically, leading to significant speedup of numerical calculations. It should be kept in mind that while in reality the galactic magnetic fields probably do drop off smoothly over some finite distance at the boundary of the diffusion zone, the particular choices of $d$ and the analytic form of $K(z)$ that we make are not physically motivated. We will show that the results of our analysis are approximately independent of these choices.

\section{Dark Matter and Galactic Propagation Models}
\label{sec:models}

\begin{table}[t]
\begin{center}
\begin{tabular}{|c|r|r|r|r|} \hline
Model & $\alpha$ & $\beta$ & $\gamma$ & $r_s$(kpc)\\
\hline
Cored isothermal &2 &2 &0  &5 \\
NFW &1 &3 &1 &20 \\
Moore &1.5 &3  &1.3  &30 \\
\hline
\end{tabular}
\caption{Dark matter density distribution profiles.}
\label{tab:halo}
\end{center}
\end{table}

The dark matter density distribution in the Milky Way halo is modeled with the generic profile
\begin{equation}
\rho(r)=\rho_\odot\left(\frac{r_\odot}{r}\right)^\gamma\left(\frac{1+(r_\odot/r_s)^\alpha}{1+(r/r_s)^\alpha}\right)^{(\beta-\gamma)/\alpha}\,,
\end{equation}
where $r_\odot=8.5$ kpc is the distance from the solar system to the galactic centre, and $\rho_\odot=0.3$ GeV cm$^{-3}$ is the local dark matter density in the solar neighborhood. We use three profiles: isothermal, Moore, and Navarro, Frenk and White (NFW) (see Table~\ref{tab:halo}). 
Since the numerically derived NFW and Moore profiles diverge at the center of the galaxy, the profile inside r$<r_0$ is replaced by the smoother profile, as in Ref.~\cite{Bessel}: 
\begin{equation}
\rho(r)=\rho_0(1+a_1\sinh(\pi x)+a_2\sinh(2\pi x) )^{1/2} \,,
\end{equation} 
where $x=r/r_0$, $\rho_0=\rho(r_0)$, 
$a_1=a_2+2\gamma$, and $a_2=8\gamma(\pi^2-9+6\gamma))/(9(3-2\gamma))$.
This renormalized profile has a continuous first derivative at $r_0$ and preserves the total number of annihilations within the core.

The positron injection spectrum $f(E)$ depends on the microscopic model of dark matter, in particular the WIMP mass and its annihilation channels or decay pattern. Motivated by PAMELA and FERMI data, we consider a rather heavy WIMP, $m_\chi = 3$ TeV (see, {\it e.g.}, Ref.~\cite{Patrick}). Since current data favor leptophilic models ({\it i.e.} those where dark matter annihilations or decays result mostly in all-leptonic final states), we focus on the following four annihilation channels:

\begin{enumerate}

\item $\chi\chi\rightarrow e^+e^-$. This channel is not favored by PAMELA and FERMI data~\cite{Patrick}, but produces a simple monochromatic injection spectrum and hence serves as a useful limit.  

\item $\chi\chi\rightarrow\mu^+\mu^-$, with the muons then decaying to give positrons via the familiar decay $\mu^+\rightarrow e^+\nu_e\overline{\nu}_\mu$.

\item $\chi\chi\rightarrow\phi\phi\rightarrow4e$, where $\phi$ is some intermediate particle of mass $m_\phi$ (for example, a gauge boson of an extra gauge symmetry that dark matter is charged under, as in Refs.~\cite{Sommer,Portal} and others).

\item $\chi\chi\rightarrow\phi\phi\rightarrow4\mu$. The motivation is the same as in 3, but the positron injection spectrum is softer in this case. 

\end{enumerate}

It is straightforward to obtain $f(E)$ for each of these models.
We consider the same final states for the case of decaying dark matter, and in fact the only difference is that the energy of the primaries ($e$ in process 1, $\mu$ in process 2, $\phi$ particles in processes 3 and 4) is halved compared to the annihilating DM scenario. 

For positron propagation in the galaxy we use the M2 and MED propagation models~\cite{Bessel} (see Table~\ref{tab:galprop}), which are compatible with cosmic ray data. We expect the enhancement of the positron flux at the solar position to be most significant in models with 
small $L$, since this both maximizes the amount of dark matter outside the diffusion zone and minimizes the energy losses of the positrons coming from outside the diffusion zone. Because of this, we do not study other well-known galactic propagation models such as M1 or MAX, in which $L>10$ kpc and no significant flux enhancement is expected.

\begin{table}[t]
\begin{center}
\begin{tabular}{|c|r|r|r|}
\hline
Model & $\delta$ & $K_0$ (kpc$^2$/Myr) & $L$ (kpc)\\
\hline
MED &0.70 &0.0112 &4\\
M2 &0.55 &0.00595 &1\\
\hline
\end{tabular}
\caption{Galactic propagation models.}
\label{tab:galprop}
\end{center}
\end{table}

To complete the description of galactic propagation in the extended formalism of Section~\ref{sec:newform}, we need to specify the parameters $D$, $K_1$ and $d$; these, and some other numerical issues, are discussed in the Appendix. 

\section{Results: Positron Fluxes}

We analyze the two galactic propagation models, three galactic halo profiles, and four dark matter annihilation scenarios specified in Section~\ref{sec:models}. In each model, we computed the halo function~\leqn{hfn_new} using the extended formalism of Section~\ref{sec:newform}, which was then used to compute positron density at the solar position ($r_\odot=8.5$ kpc, $z_\odot=0$) as a function of energy. For comparison, we also computed the halo function and positron density within the conventional formalism (Sec.~\ref{sec:oldform}), which neglects the contribution due to dark matter annihilations outside of the diffusion zone. The positron flux measured by a Solar System based experiment is given by
\beq
\Phi_{e^+}(E) \,=\, \frac{\beta_{e^+}}{4\pi} \,\psi (r_\odot, z_\odot, E)\
\eeq{flux}
where $\beta_{e^+}$ is the velocity of a positron of energy $E$. 

\begin{figure}[t]
\centering
\begin{tabular}{ccc}
\includegraphics[width=2.1in,height=1.6in]{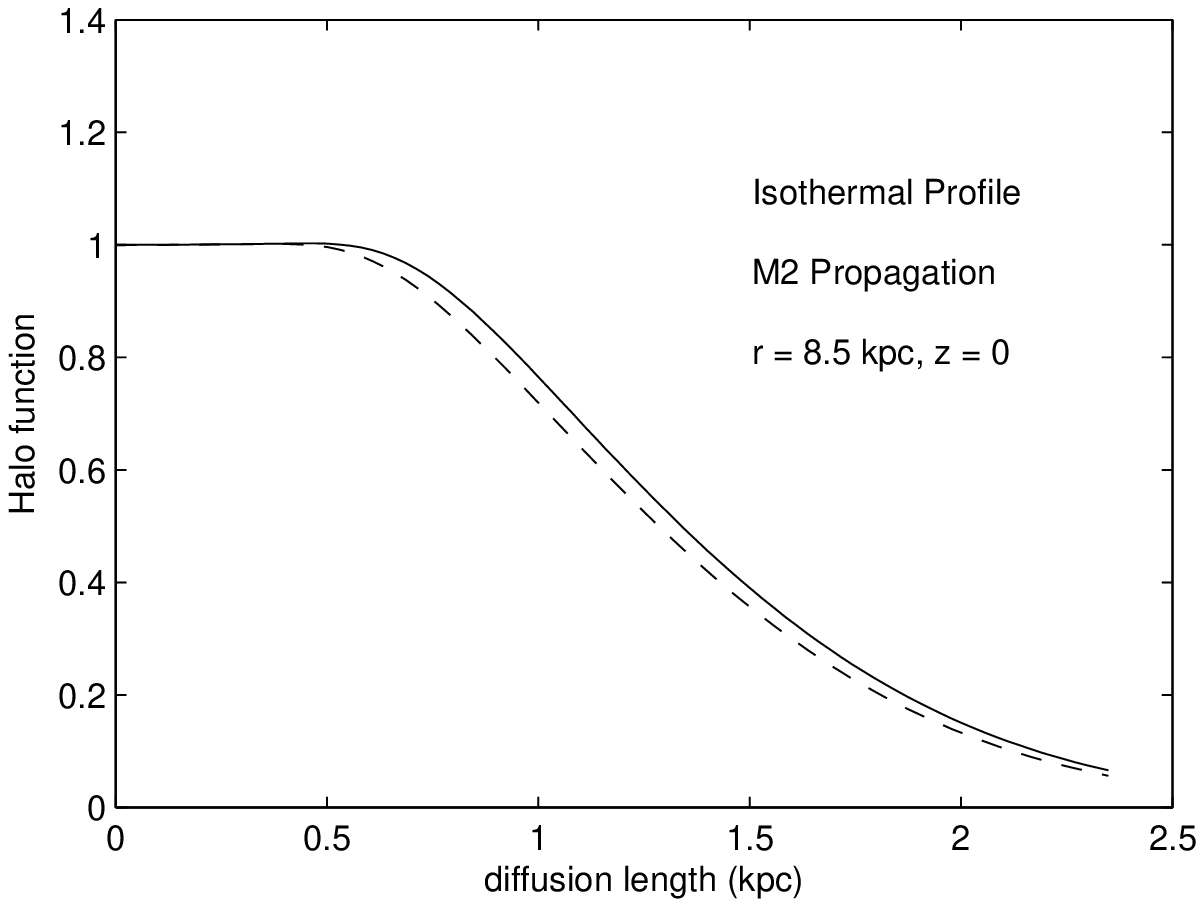}&
\includegraphics[width=2.1in,height=1.6in]{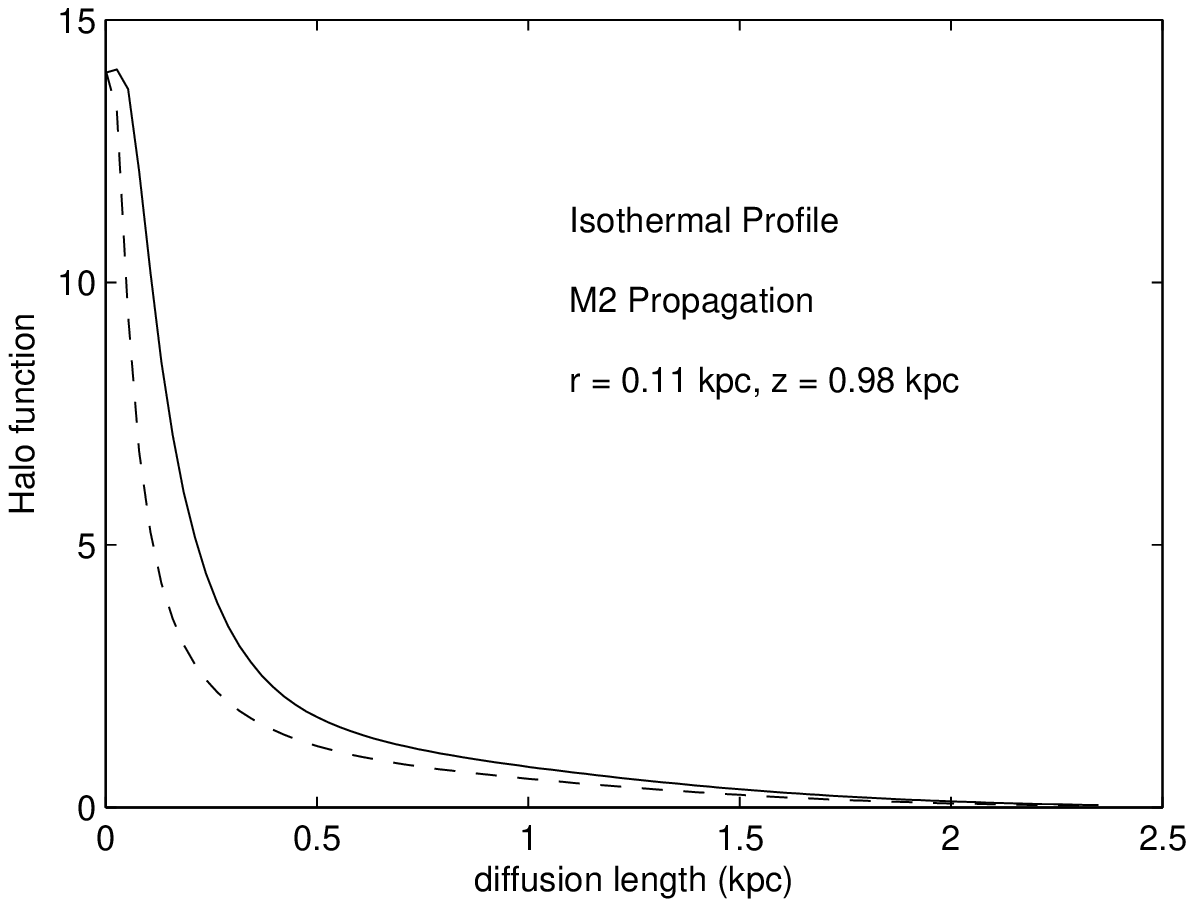}&
\includegraphics[width=2.1in,height=1.6in]{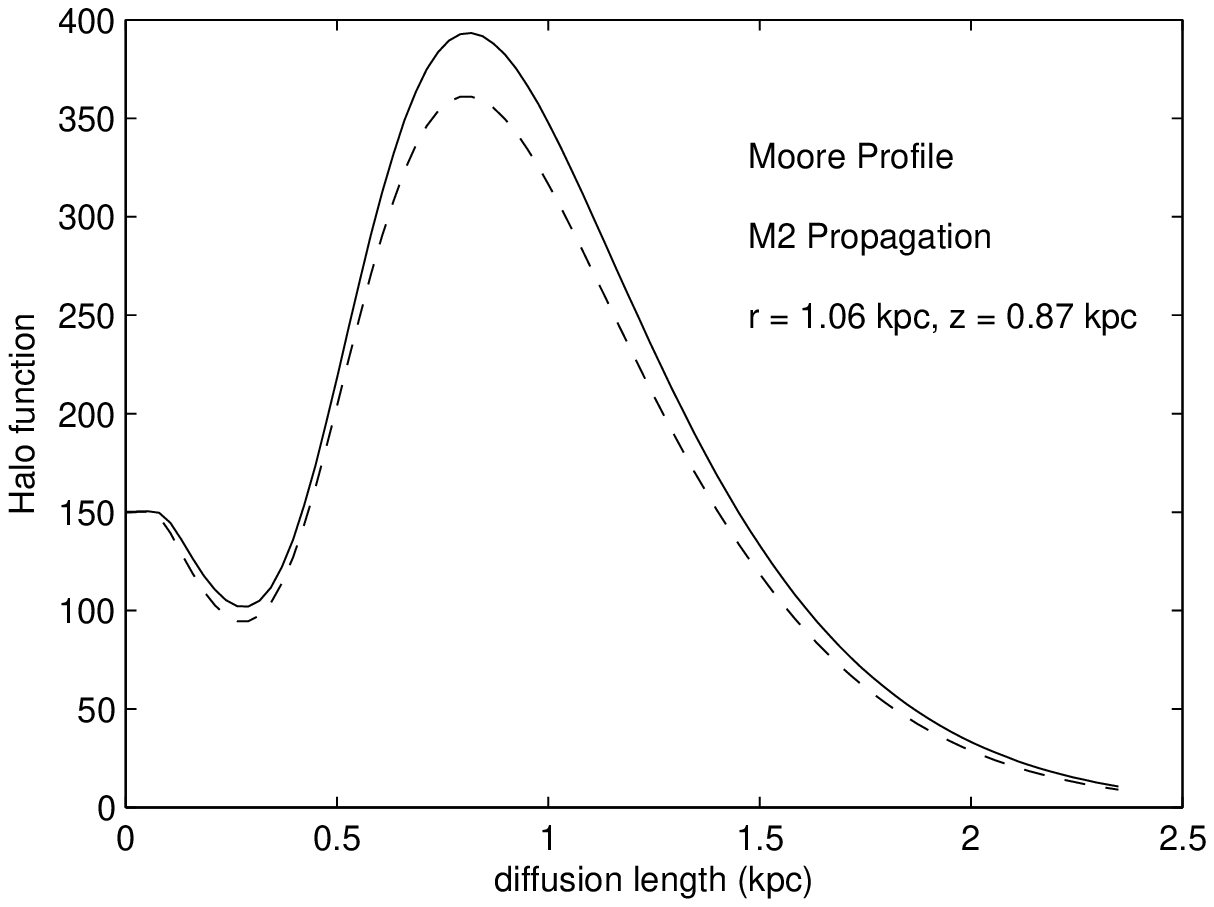}\\
\end{tabular}
\caption{Examples of conventional (dashed) and extended (solid) halo functions.}
\label{halos}
\end{figure}

Figure \ref{halos} shows the correction to the halo function for a few cases. The corrections are generally small, even for the M2 model with $L=1$ kpc. As expected, the correction is larger for flatter profiles and in regions closer to the diffusion zone boundary. 

\begin{figure}[t!]
\centering
\begin{tabular}{cc}
\includegraphics[width=3in,height=1.8in]{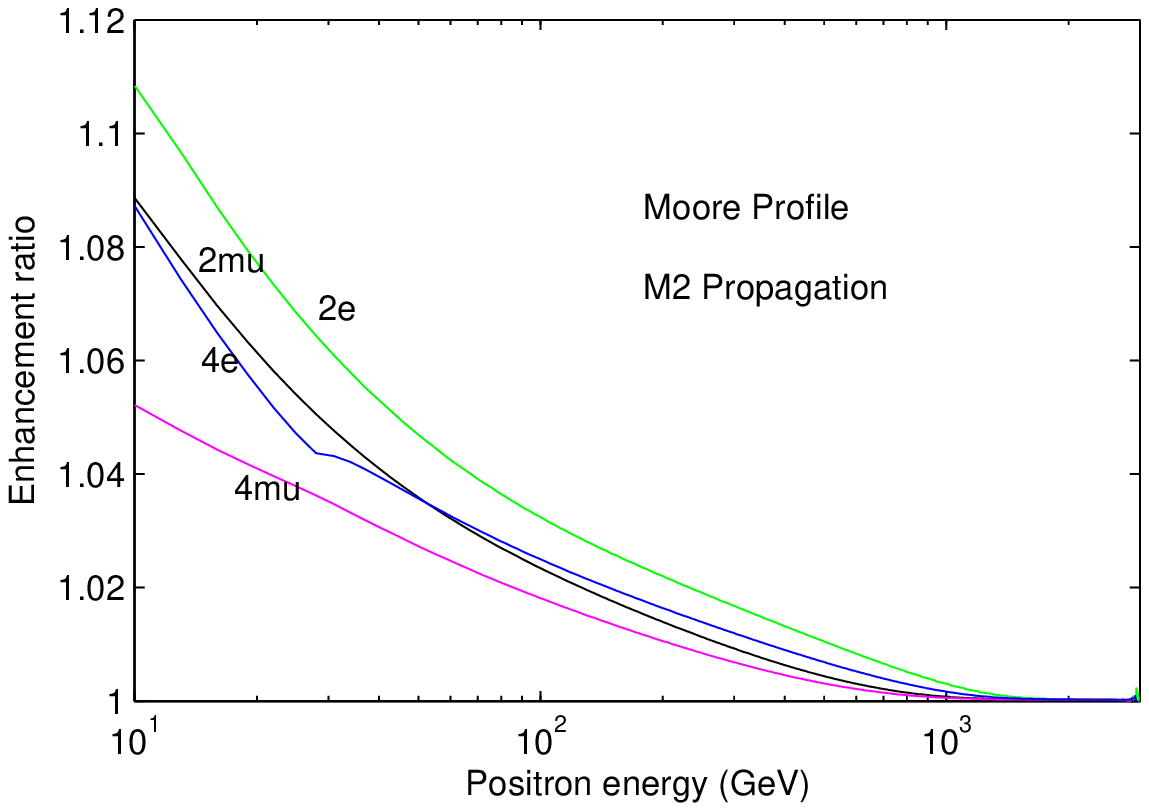}&
\includegraphics[width=3in,height=1.8in]{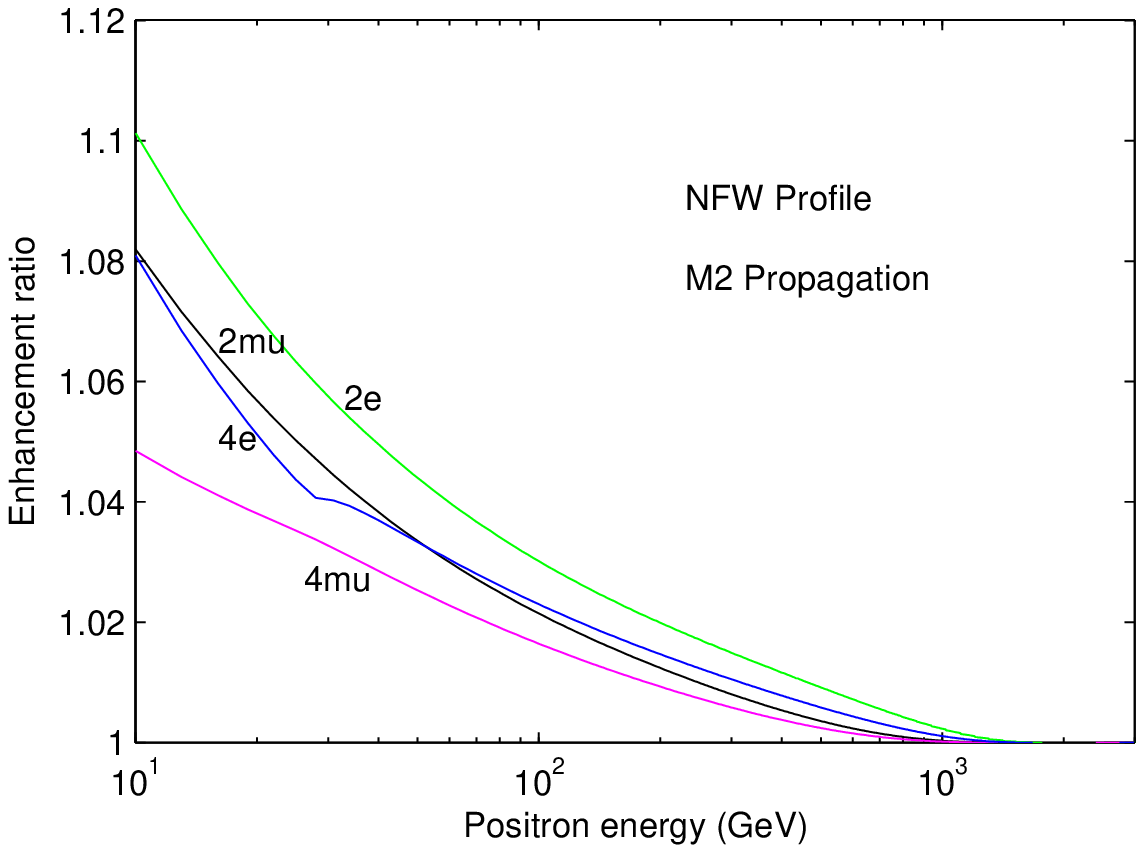}\\
\includegraphics[width=3in,height=1.8in]{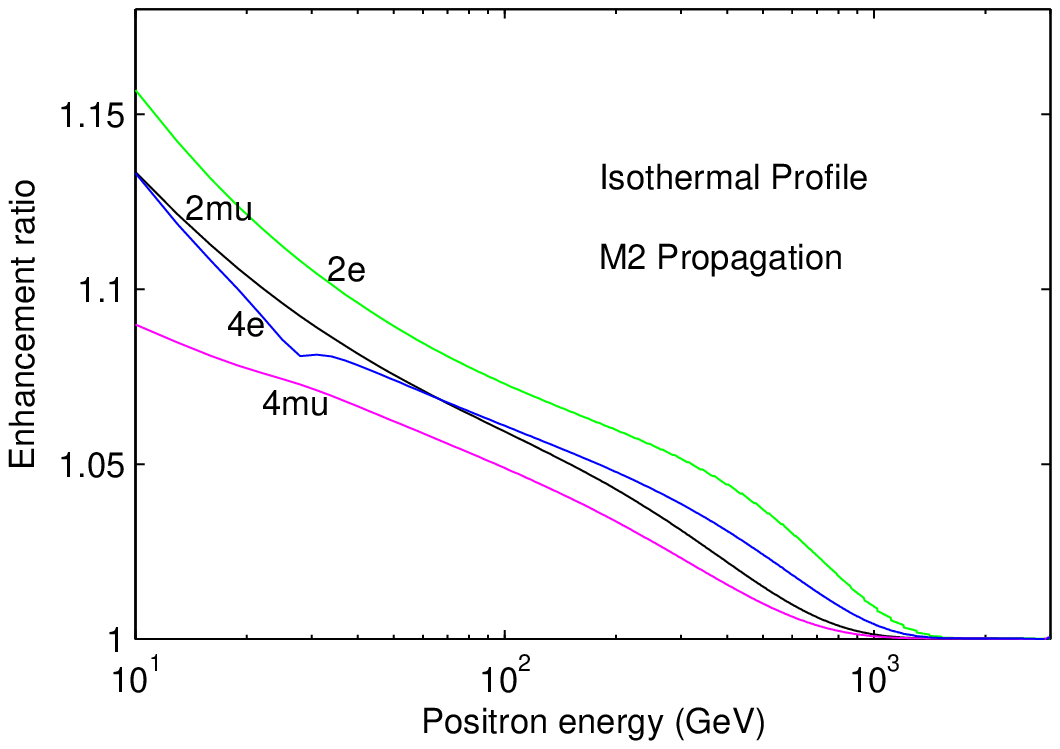}&
\includegraphics[width=3in,height=1.8in]{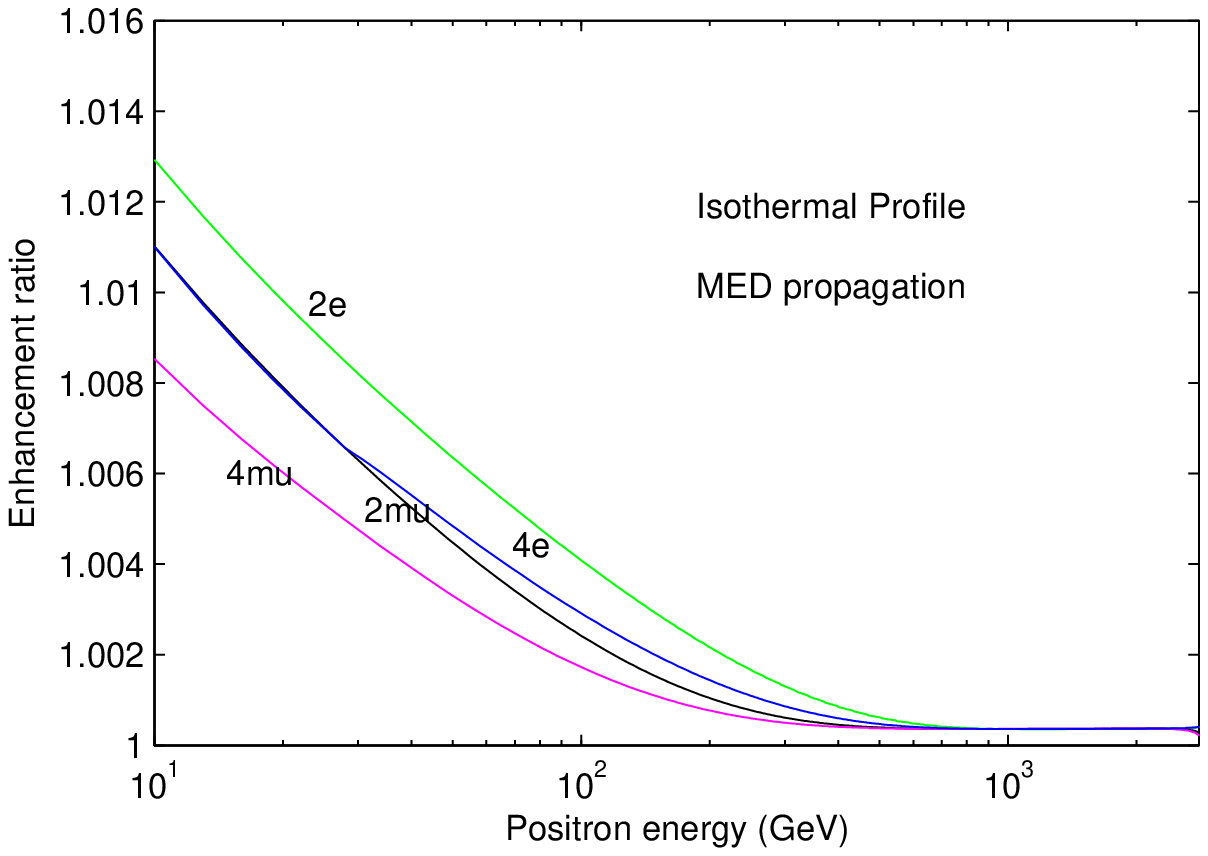}\\
\end{tabular}
\caption{Positron flux enhancement for annihilating dark matter, for $m_{\chi}$ = 3 TeV.}
\label{enhancements}
\end{figure}

In Fig.~\ref{enhancements}, we plot the ratios of the full positron flux and the flux computed within the conventional formalism for various scenarios. Overall, the enhancements we observe are not large, typically in a few-\% range, and are thus smaller than other astrophysical uncertainties at this point. 
The plots in Fig.~\ref{enhancements} demonstrate many expected features. First, the enhancement is largest for annihilation into $e^+e^-$, which has the most energetic input spectrum, and progressively decreases for less energetic input spectra. Second, the plots are consistent with the notion that for M2 propagation the spectrum at Earth is influenced largely by sources within a few kpc and, in particular, is insensitive to the large cusp at the center of the galaxy. In particular, although the Moore profile is more cusped at the galactic center than the NFW profile, in the solar neighborhood the dark matter density drops off with $\textit{z}$ faster in the NFW profile than in the Moore profile. Consistent with this, the flux enhancement is smaller in the NFW profile. Third, the enhancement is larger at lower positron energies, consistent with the fact that positrons lose energy as they propagate through the intergalactic medium and hence energetic positrons entering from the halo outside the diffusion zone will arrive at the Earth at lower energies. Fourth, the enhancement is almost negligible for MED propagation, since there is negligible amount of dark matter outside the diffusion zone compared to inside it in this case, and the diffusion zone boundary is also farther away from the Earth. 
All these features confirm that the extended formalism we suggest, and our numerical approximations, are physically sensible.

\subsection{Decaying Dark Matter}

Dark matter decaying with a lifetime of about 10$^9$ times longer than the age of the universe can also explain the PAMELA and FERMI excesses~\cite{DecayDM}. While dark matter annihilation rates are proportional to $\rho_{\chi}^2$, the rate for decaying dark matter is proportional to $\rho_{\chi}$ instead, which implies that the relative contribution from the halo exterior to the diffusion zone should be greater. 

\begin{figure}[h]
\centering
\begin{tabular}{cc}
\includegraphics[width=3in,height=1.8in]{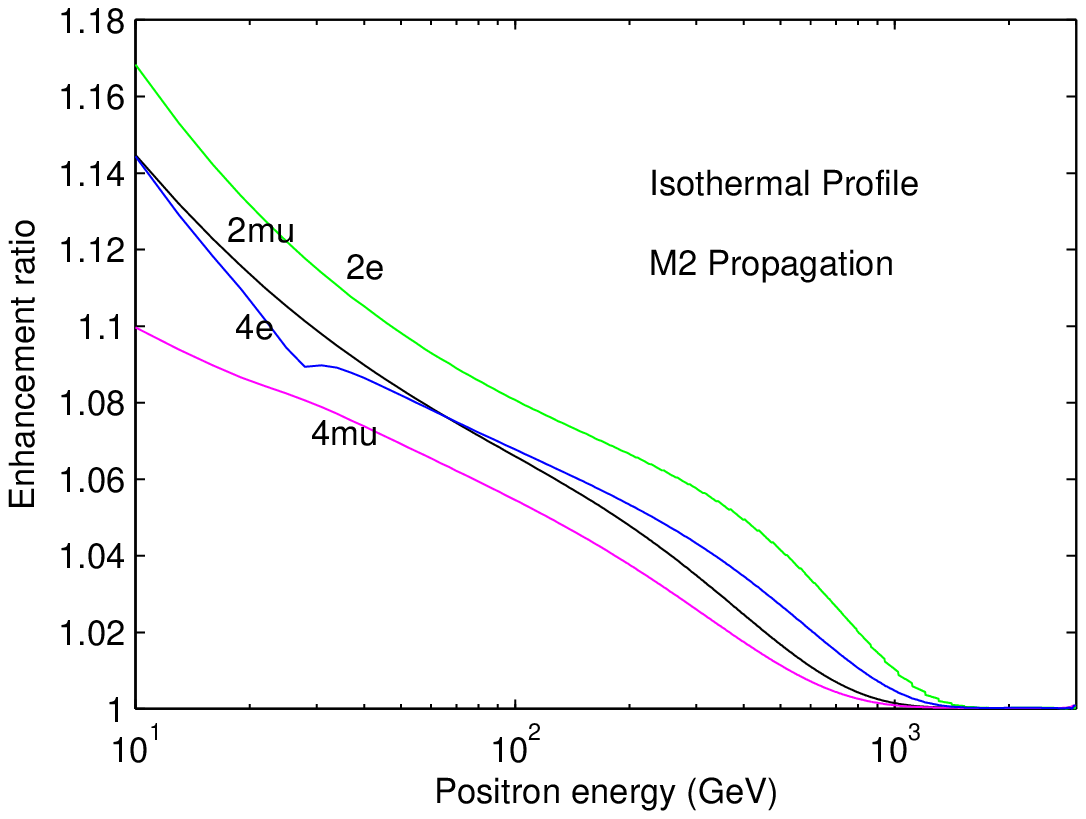}&
\includegraphics[width=3in,height=1.8in]{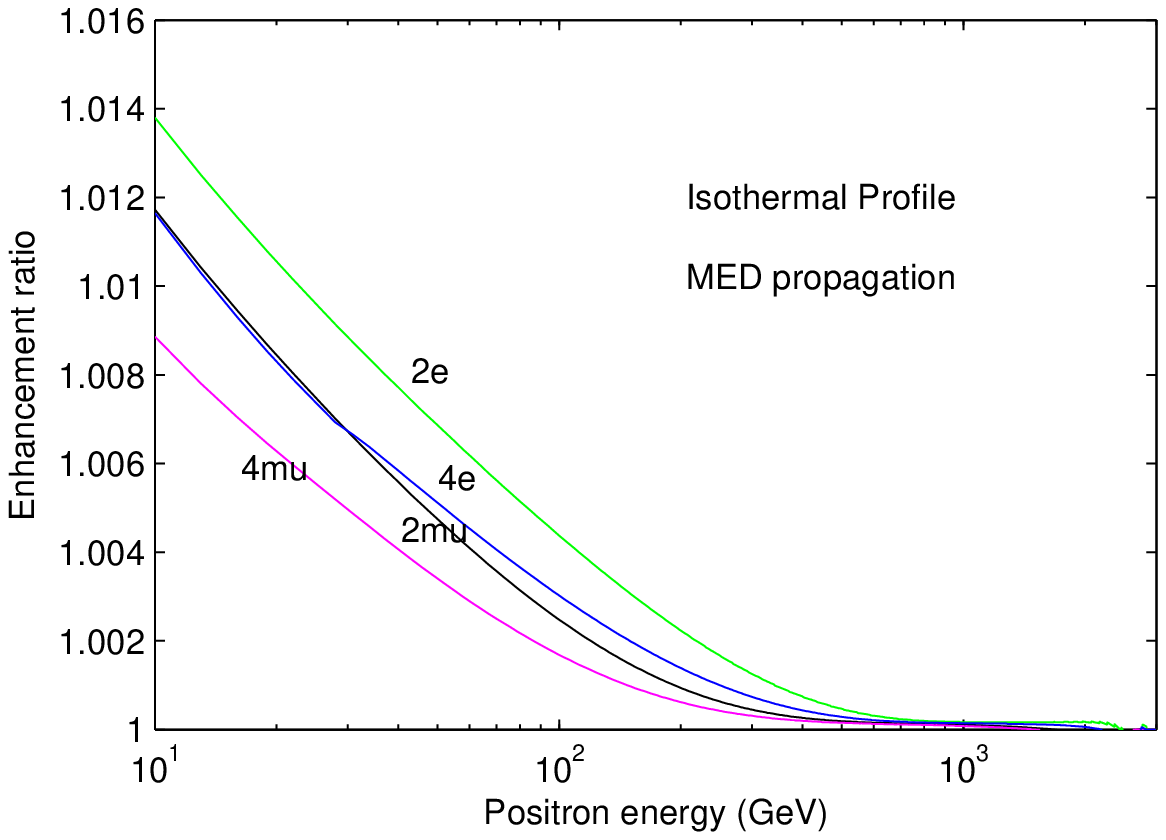}\\
\end{tabular}
\caption{Positron flux enhancement for decaying dark matter, for m$_{\chi}$ = 6 TeV.}
\label{enhancements_dec}
\end{figure}

We computed the enhancement for decaying dark matter for the isothermal profile since, for a given galactic propagation model, the enhancements are the largest for this profile. The results are plotted in Fig.~\ref{enhancements_dec} and should be contrasted with the bottom row of Fig.~\ref{enhancements}. For MED propagation the enhancement remains negligible; for M2 propagation the increase over the annihilating dark matter scenario is very small because the region of influence only extends to a few kpc, where differences between $\langle\rho_{\chi}^2\rangle$ and $\langle\rho_{\chi}\rangle$ are small.

\section{Gamma Rays from Inverse Compton Scattering}
\label{sec:ICSresults}

The positron flux corrections calculated in the previous section are at the solar position, which lies at $z=0$ and hence only gets minimal corrections. The more significant corrections occur close to the diffusion zone boundary, but these cannot be observed directly. On the other hand, gamma ray flux from inverse Compton scattering (ICS) off positrons is sensitive to positron density throughout the galaxy, since a photon scattering off an energetic positron towards the Earth anywhere in the galaxy will travel unperturbed through the interstellar medium and can be detected. Hence large corrections to positron density close to the boundary can significantly affect the ICS spectrum measured at Earth. 

A semi-analytic calculation of the ICS energy spectrum from dark matter annihilation is presented in~\cite{Patrick} and~\cite{ICS}. The ICS flux is expressed (see Eq.(10) of~\cite{Patrick}) as
\beq
\frac{d\Phi_{\gamma'}}{dE_{\gamma'}}=\sum_i H_{iIC}\frac{9r_\odot \langle\sigma v\rangle}{64\pi\langle E_{\gamma i}\rangle}\left(\frac{\rho_\odot}{m_\chi}\right)^2\,.
\eeq{ics1}
Here we have combined the two separate dimensionless parameters $J_{IC}$ and $G_{IC}$ defined in 
Eq.~(11) of Ref.~\cite{Patrick} into a single parameter
\beq
H_{iIC}=m_e^4\int d\Omega\int_{l.o.s}\frac{ds}{r_\odot}\frac{u_{\gamma i}}{u_{tot}}\int\frac{dE_\gamma}{E_\gamma}f_{\gamma i}(E_\gamma)\int\frac{dE_e}{E_e^4}\frac{f_{IC}}{R(E_e)}\int^{m_\chi}_{E_e} dE'f(E')\tilde{I}(E_e,E',r)\,.
\eeq{ics2}
This is necessitated by the explicit position dependence of the halo function $\tilde{I}$ in our treatment, which requires it to be integrated over both position and energies. $E_\gamma$ and $E_{\gamma'}$ denote photon energies before and after scattering. The sum over $i$ accounts for the three components of galactic light that can scatter off energetic positrons: CMB, starlight, and starlight rescattered by dust.  The energy density profiles $u_i$, energy spectra $f_{\gamma i}$, mean energies $\langle E_{\gamma i}\rangle$, the inverse Compton factor $f_{IC}$ in the scattered photon spectrum, and relativistic correction $R(E_e$) are as presented in Ref.~\cite{Patrick}. 
Eq.~\leqn{ics2} shows that if the line of sight ends at the diffusion zone boundary, the halo function $\tilde{I}$ at that position, which receives significant corrections from the extended halo, directly enters the calculation for the ICS flux. 
 
\begin{figure}[t!]
\centering
\begin{tabular}{cc}
\includegraphics[width=3in,height=1.8in]{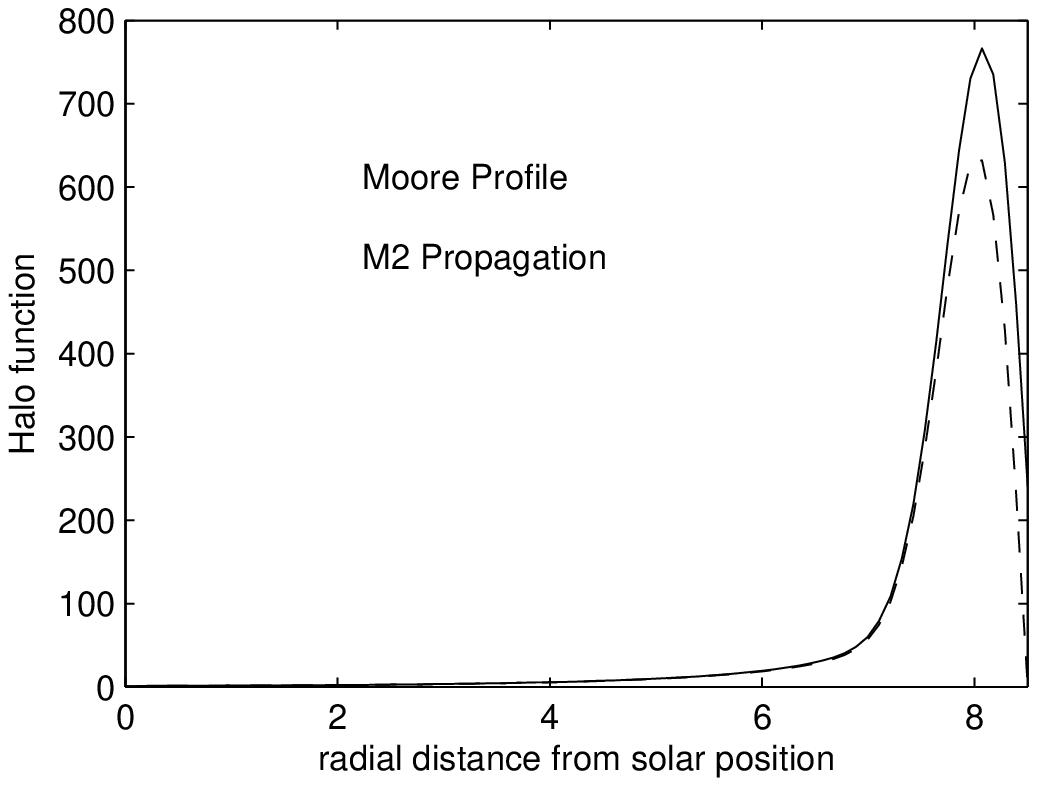}&
\includegraphics[width=3in,height=1.8in]{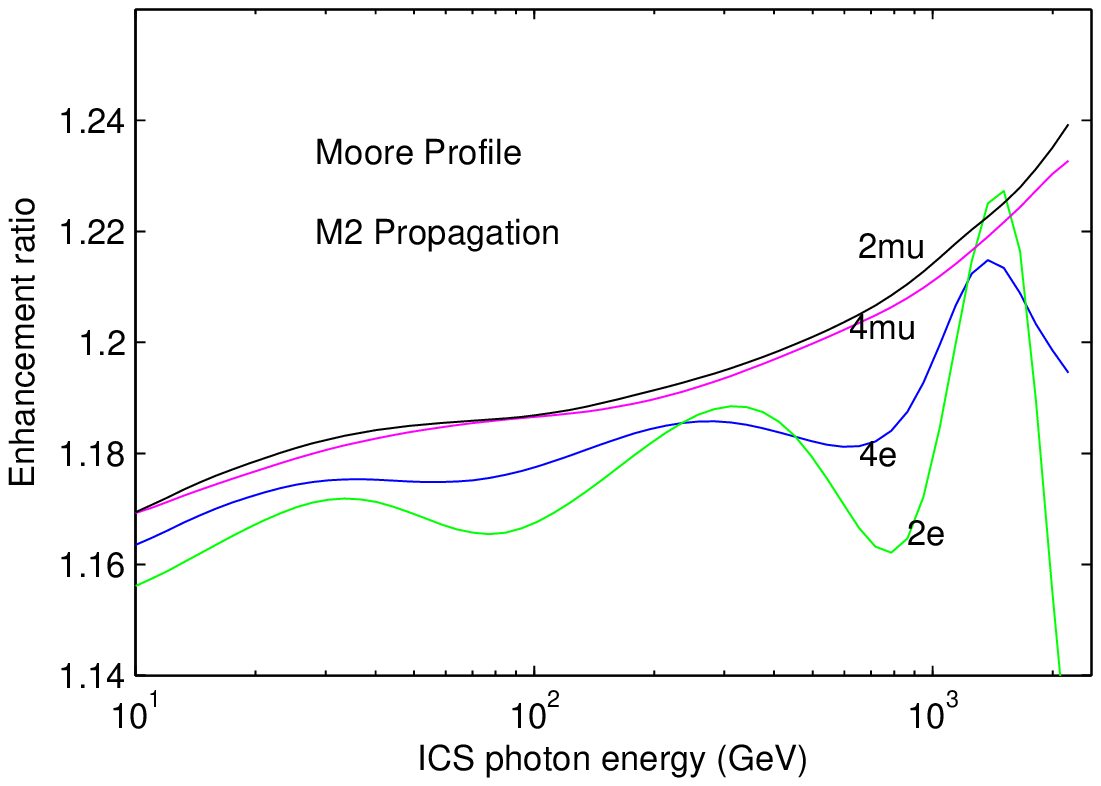}\\
\includegraphics[width=3in,height=1.8in]{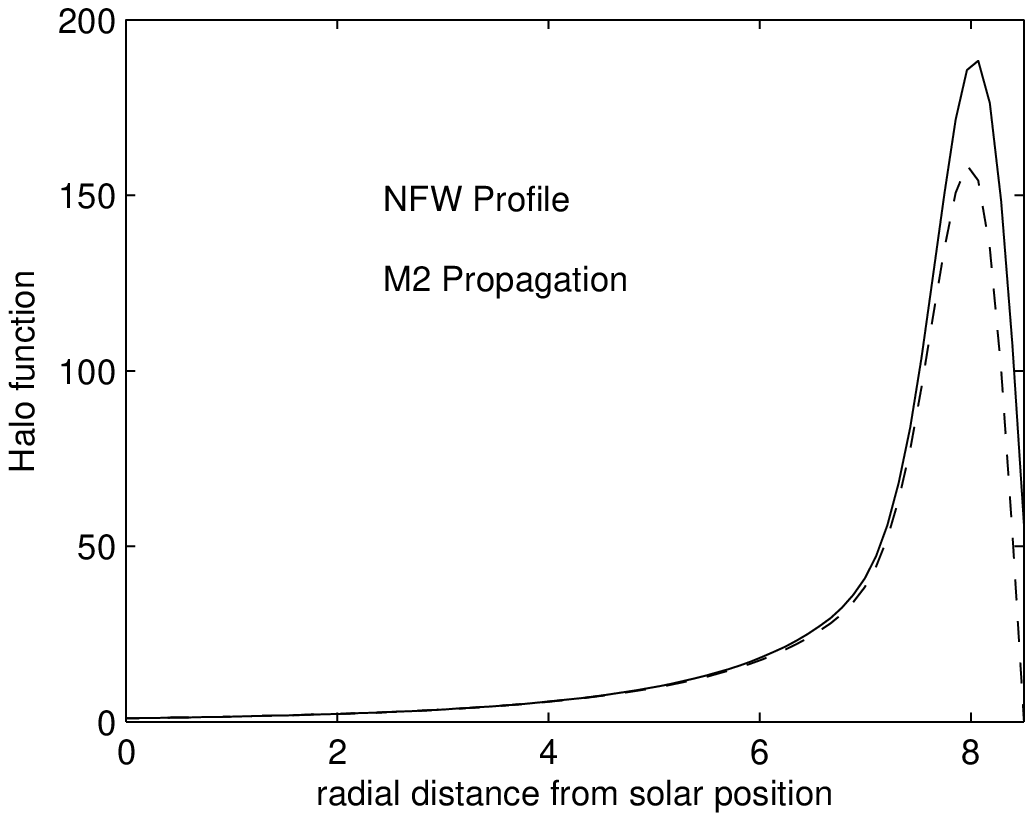}&
\includegraphics[width=3in,height=1.8in]{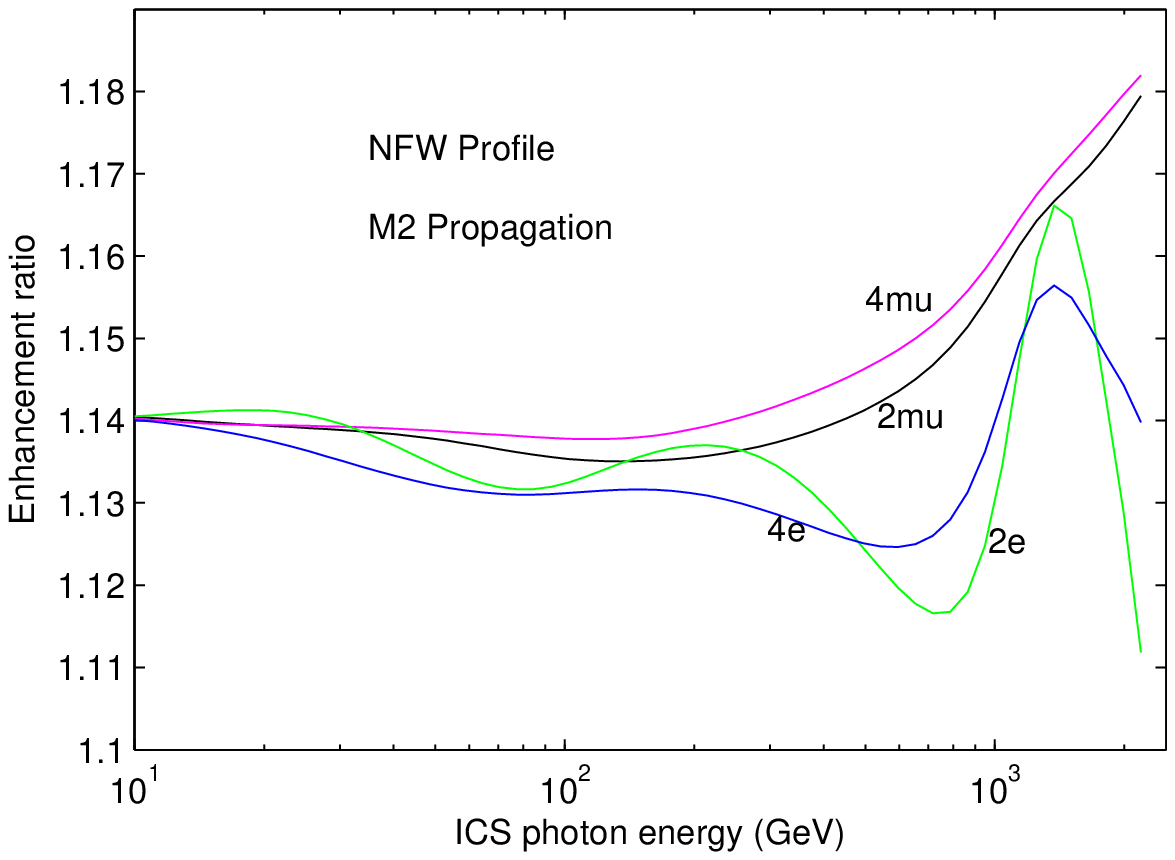}\\
\includegraphics[width=3in,height=1.8in]{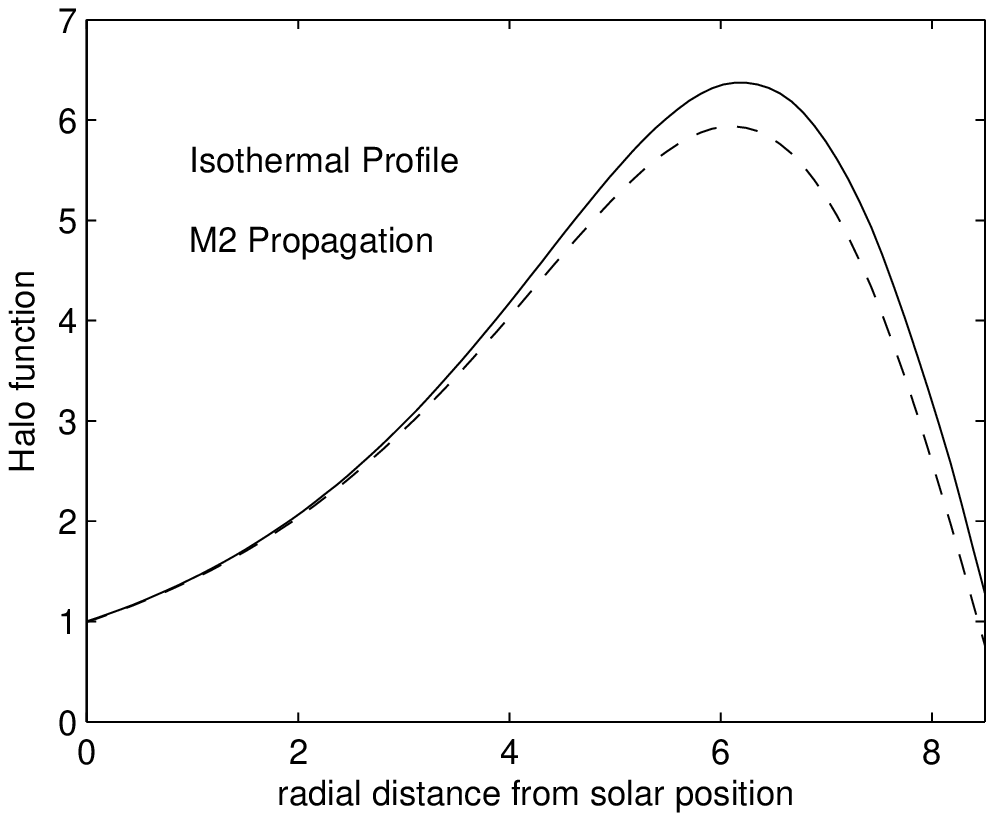}&
\includegraphics[width=3in,height=1.8in]{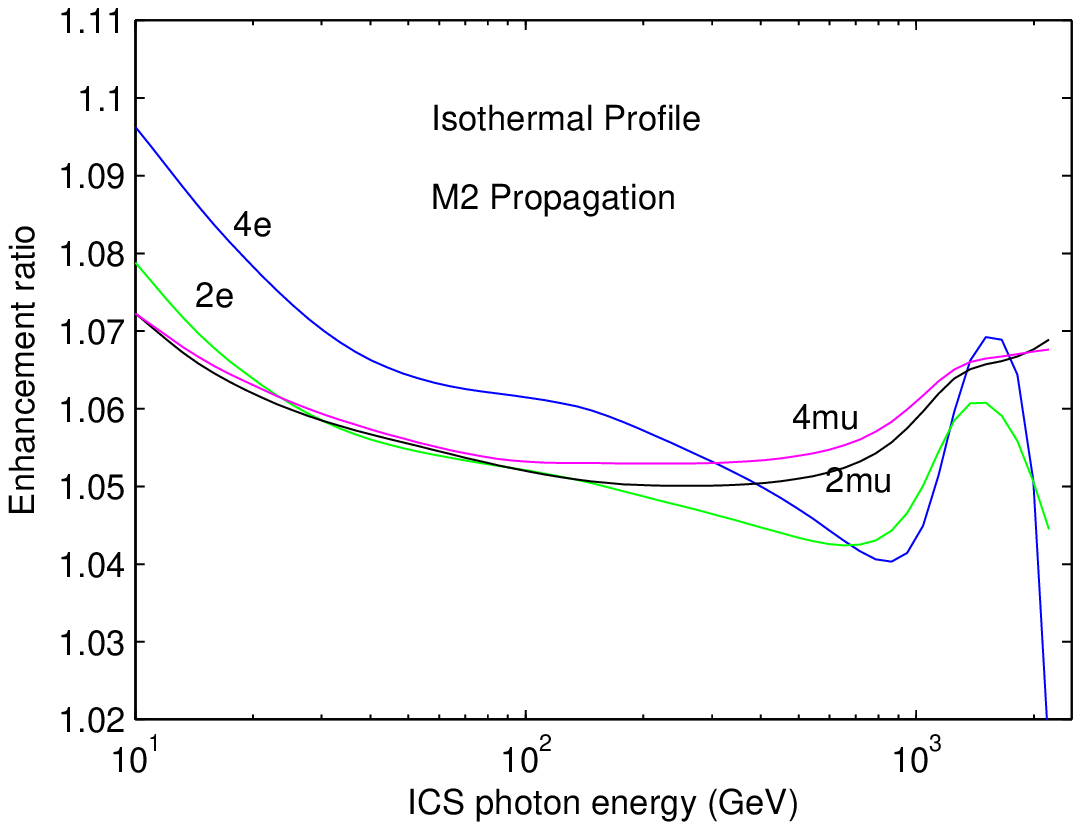}\\
\end{tabular}
\caption{Left column: Halo function from conventional (dashed) and extended (solid) formalisms as a function of position along line of sight for $\lambda_D=$ 0.5 kpc. Right column: ICS gamma ray flux enhancement for annihilating dark matter, $m_\chi$ = 3 TeV.}
\label{ICSflux}
\end{figure} 
 
As an example, we consider the line of sight from the Earth ($r=8.5$ kpc, $z=0$) to the edge of the diffusion zone closest to the galactic center ($r=0$, $z=1$ kpc in the M2 model). In this case, the largest contribution to the ICS flux enhancement comes from regions close to the diffusion zone boundary: the photon density is the greatest there because of the proximity to the galactic center, and the positron density is zero or close to zero in the conventional formalism because of the boundary conditions but can be significant in the extended formalism (see Fig.~\ref{ICSflux}, left column). The second column in Fig.~\ref{ICSflux} shows the enhancement in the ICS flux, calculated using Eq.~\leqn{ics1} and Eq.~\leqn{ics2}. As expected, the enhancement in the ICS gamma ray flux is more significant than in positron flux. While there is essentially no correction to the positron flux around the TeV scale, the ICS flux at this energy receives corrections of about 20\%. 
Note that the enhancement plots -- especially in the $e^+e^-$ channel -- show three distinct ``bumps". These correspond to the three different galactic gamma-ray components: CMB, starlight, and starlight rescattered off dust.

\section{Conclusions}
We summarize our findings as follows:
\begin{itemize}

\item Including contributions from the extended dark matter halo to positron density in the galactic diffusion zone coming from dark matter annihilation or decay can result in corrections of up to 17\% in the observed positron flux and up to 24\% in the ICS gamma ray flux. 

\item For positron flux, the enhancement is progressively lower at higher energies since positrons coming in from the halo lose energy in the diffusion zone and arrive at detectors with lowered energies.  While we see enhancements of up to 17\% at 10 GeV, enhancements are $<$10\% for all considered cases for energies $>$100 GeV. 

\item For ICS gamma ray flux, the bulk of the correction comes from photons scattering off energetic positrons close to the diffusion zone boundary, hence the enhancement is maintained even at the highest photon energies. 

\item The enhancements are most significant for the M2 propagation model, where the diffusion zone is only 2 kpc thick and hence the bulk of the dark matter halo lies outside this zone. For MED model (and presumably other models with larger $L$), the corrections are negligible. 

\item At present, experimental uncertainties on flux measurements as well as astrophysical uncertainties in the positron and ICS fluxes -- these come from numerous sources, such as uncertainties from dark matter profiles, propagation models, energy density and spectra of photons in the galaxy, and the simplifications made to the transport equation to describe positron propagation -- remain significantly greater than the additional contribution from the dark matter halo beyond the diffusion zone. It does not need to be included in fits to data at this stage, but should be considered when accuracy to better than 25\% is needed.

\end{itemize}

The extended formalism can also be augmented in a straightforward manner to include contributions from other important sources, such as dwarf galaxies, that are impossible to incorporate in the conventional formalism.

\vskip0.5cm

\noindent{\large \bf Acknowledgments} 
\vskip0.1cm

We are grateful to Peter Graham and Patrick Meade for useful discussions. This research is supported by the U.S. National Science Foundation through grant PHY-0757868 and CAREER award PHY-0844667. 

\appendix

\section{Comments on Numerical Issues}

This appendix discusses some details of our numerical calculations.

{\it Choice of parameters ---}
Parameter choices were dictated by the need to carry out all computations in a reasonable amount of time. For each dark matter density profile, \textit{D} was chosen such that $\rho^2_\chi$ at ($r=0$, $z=D$) is 5\% of the corresponding value at the diffusion zone boundary ($r=0$, $z=L$). \textit{K$_1$} was typically chosen to be 2000 times \textit{K$_0$}, while we typically used $\textit{d}=0.2$ kpc and $\textit{d}=0.5$ kpc for M2 and MED propagation respectively.  Although $\textit{R} = 20$ kpc is the standard radius of the diffusion zone cylinder, we used $\textit{R} = 11$ kpc for runs with NFW and Moore profiles to save computation time. This choice is justified because $(i)$ there is negligible amount of dark matter beyond $\textit{R} = 11$ kpc for these profiles, and $(ii)$ all our results for these profiles use the M2 propagation model, where the diffusion zone cylinder height is so small that positron abundance is determined primarily by losses in the vertical (z) direction, and the diffusion region forgone by using the smaller radius does not significantly affect the results.

\begin{figure}[h!]
\centering
\begin{tabular}{cc}
\includegraphics[width=3in,height=1.8in]{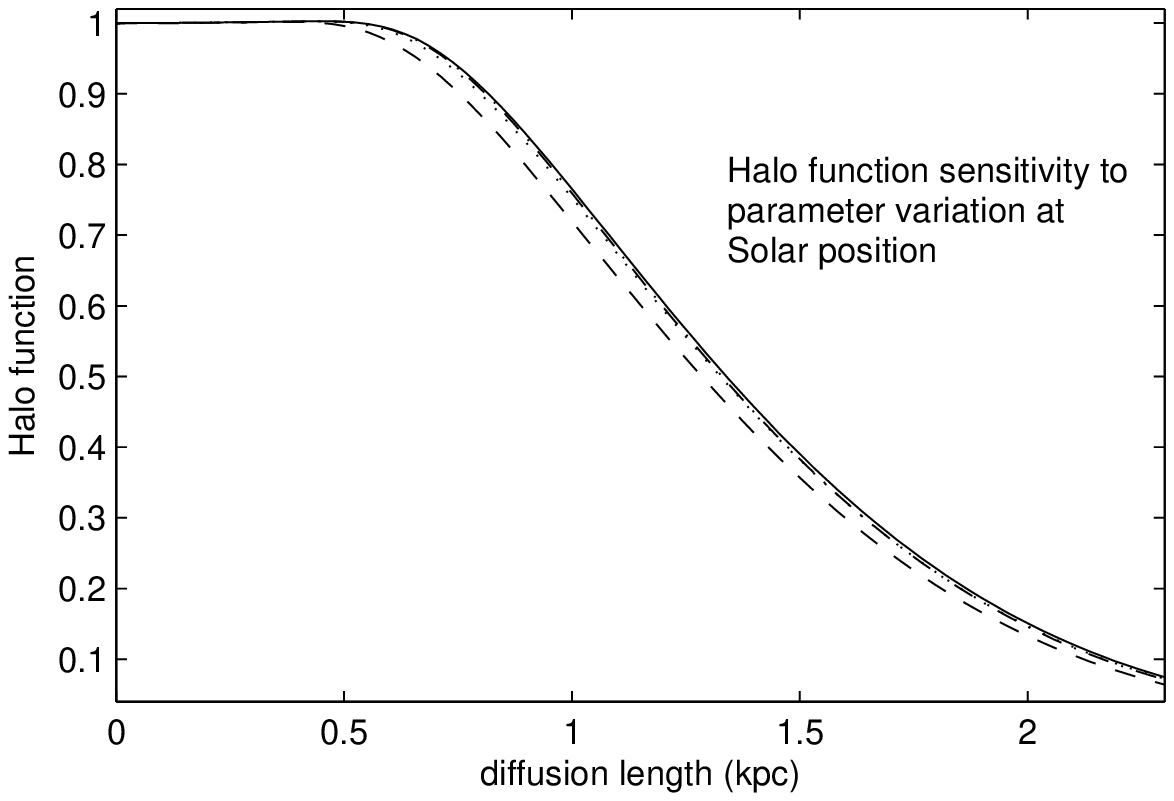}&
\includegraphics[width=3in,height=1.8in]{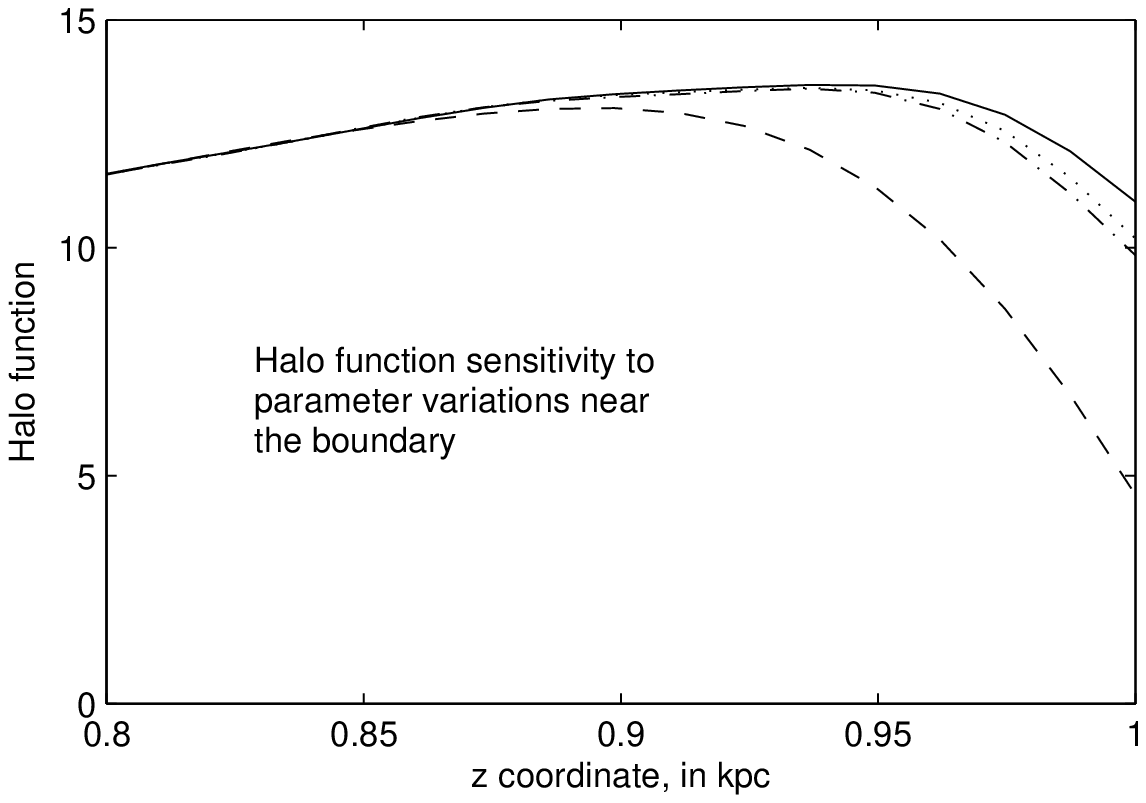}\\
\includegraphics[width=3in,height=1.8in]{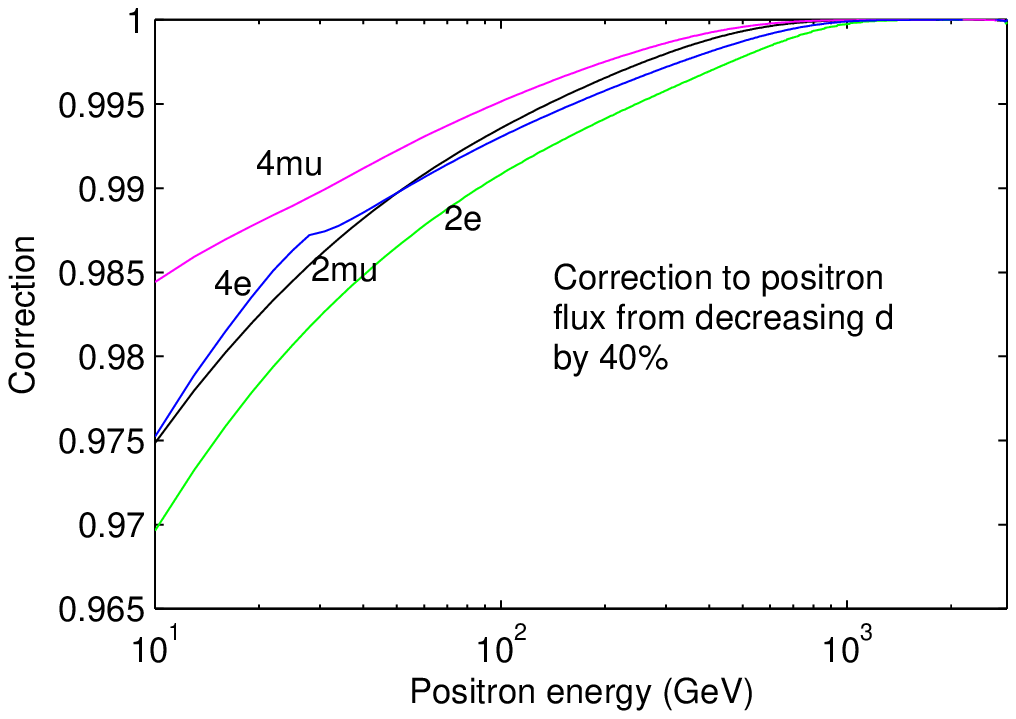}&
\includegraphics[width=3in,height=1.8in]{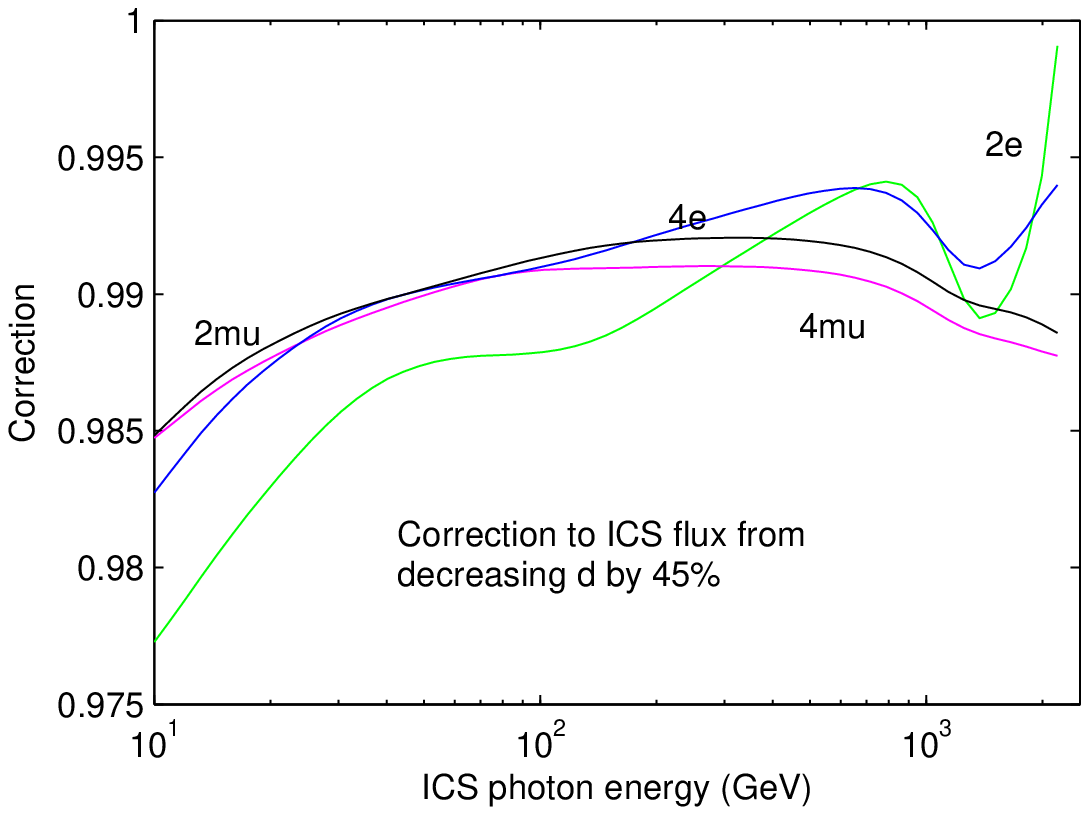}\\
\includegraphics[width=3in,height=1.8in]{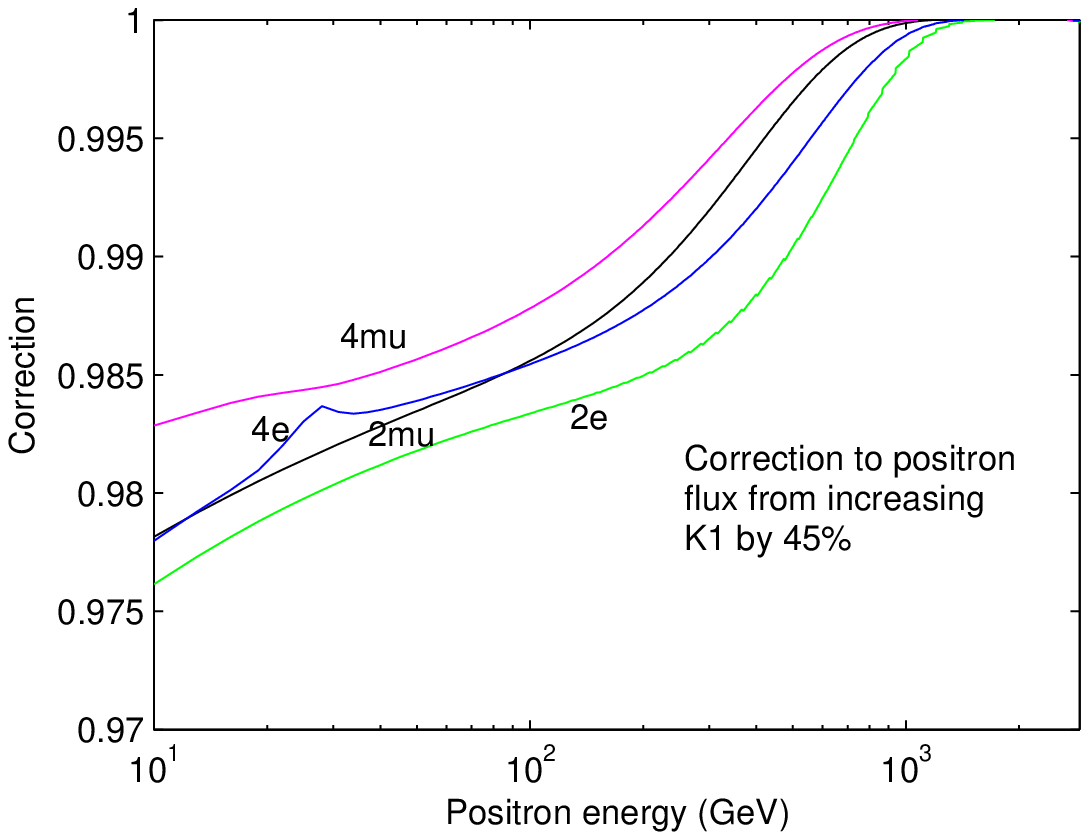}&
\includegraphics[width=3in,height=1.8in]{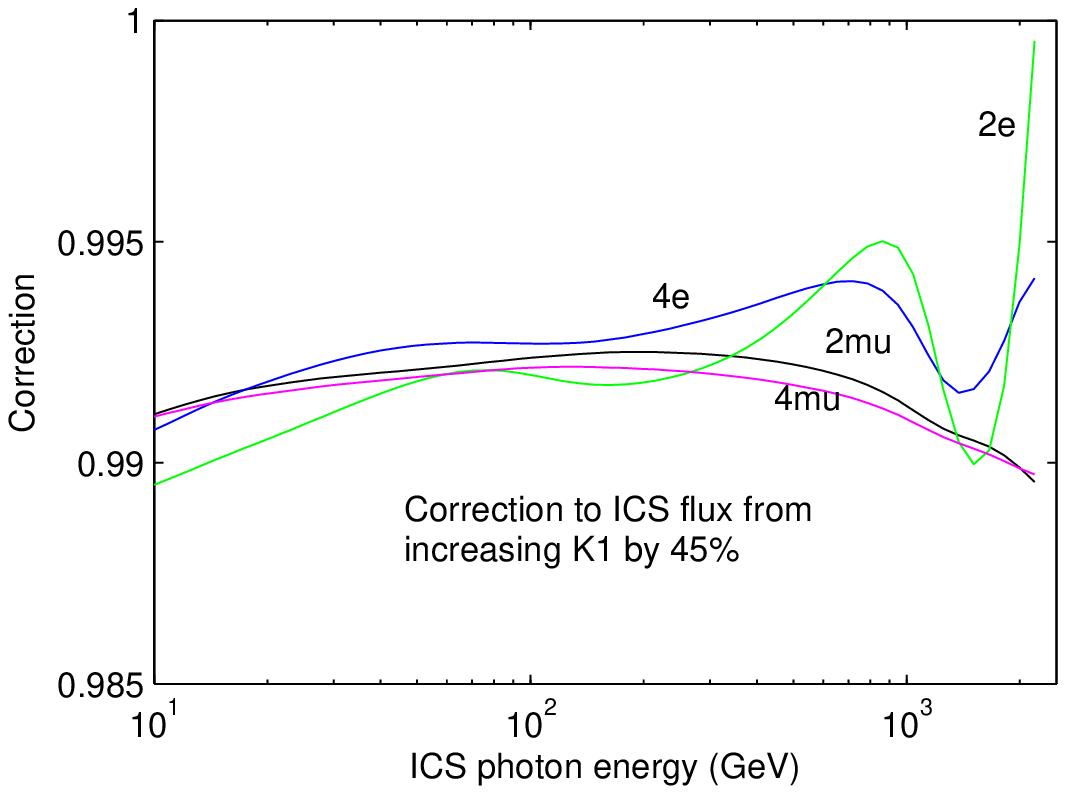}\\
\end{tabular}
\caption{Halo function, positron flux, and ICS flux sensitivity to changes in \textit{d} and \textit{K$_1$}.}
\label{convergences}
\end{figure}

The results should not be sensitive to the choices of \textit{K$_1$} or \textit{d}. To verify this, we varied \textit{K$_1$} and \textit{d} and checked how this affected our results. This is shown in Fig.~\ref{convergences} for the isothermal profile, M2 propagation. The first column shows the effects on the positron flux at Earth. The top plot shows how the halo function changes at the solar position: the dashed and solid curves correspond to halo functions in the conventional and extended formalisms, while the two curves in between correspond to \textit{d} decreased by 40\% (dot dashed curve) and \textit{K$_1$} increased by 45\% (dotted curve) respectively. The next two plots show the corrections to the positron flux from these two parameter variations; the correction factor on the y-axis is calculated as the ratio of the flux after parameter variation to the flux before parameter variation. 
 Likewise, the second column shows the corresponding effects on the ICS spectrum. The top plot shows how the halo function changes (along the line of sight used in Section~\ref{sec:ICSresults}) near the diffusion zone boundary, where the effect should be the most important, for $\lambda_D =$ 0.08 kpc (right). The next two plots show the corrections to the ICS flux from the two aforementioned variations.  
 The sensitivity to these variations is $<$3\%, while the effects we are studying give enhancements of 10-20\% (Fig.~\ref{enhancements} and Fig.~\ref{ICSflux}). Therefore sensitivity to unphysical parameter choices is small compared to the physical correction we are studying, and does not affect out conclusions. 
 
 To get results to converge, 2500 terms were needed in the Fourier series expansion. The Bessel series expansion required 60 terms for the isothermal profile and 625 terms for the NFW and Moore profiles. 

 {\it Evaluating the halo function ---} 
 To evaluate the exponential in Eq.~\leqn{hfn_new}, we diagonalized the $\textbf{A$_i$}$ matrices. 
 In a basis \textbf{V}$_i$ in which \textbf{D}$_i$ = \textbf{V$_i^{-1}$}\textbf{A}$_i$\textbf{V}$_i$ is diagonal, Eq.~\leqn{hfn_new} becomes 
\beq
I(r,z,\eps, \eps_S) \,=\,\sum_{i}\sum_{n} J_0\left(\frac{\alpha_ir}{R}\right)\sin\left(\frac{n\pi(z+D)}{2D}\right)\
\left( \textbf{V}_i\exp \Bigl[-(t-t_S) {\textbf D}_i \Bigr] \textbf{V$_i^{-1}$}{\textbf R}_{i} \right)_n,
\eeq{hfn_new_diag}
and the exponential is straightforward to calculate. This method was found to be more efficient than Taylor-expanding the exponential. The diagonalization procedure for large \textbf{A$_i$} can be sped up by noting that its entries are largest along the diagonal and get progressively smaller away from the diagonal, so we can approximate \textbf{A$_i$} to be $n\times n$ block diagonal, hence splitting the diagonalization task into $n$ smaller, independent tasks. The $2500\times 2500$ \textbf{A$_i$} matrices were approximated to be block diagonal, containing five $500\times 500$ blocks; to minimize the error from this approximation on the edge terms of the blocks, each block was diagonalized by first diagonalizing a larger $600\times 600$ block, and throwing away 50 terms from each end. The results from this procedure were found to be in agreement with results obtained by diagonalizing the full \textbf{A$_i$} matrix for a few cases that were checked.

{\it Boundary position correction ---}
The form of the diffusion coefficient in Eq.~\leqn{Kz} has many benefits, but it also has a flaw: we ideally want $\textit{K}(z)$ to quickly deviate from \textit{K$_0$} immediately outside the boundary $z = L$, but the function in Eq.~\leqn{Kz} rises slowly at first before becoming steeper because of the way trigonometric functions are shaped. Thus, over a small distance $\epsilon$ just outside the boundary, $\textit{K}(z)\sim\textit{K$_0$}$, and this region $L\leq z\leq L+\epsilon$ traps positrons just like the diffusion zone. For a fairer comparison, it is therefore necessary to shift the diffusion zone boundary inwards by a distance $\epsilon$ (or move the boundary for the conventional calculation outwards by distance $\epsilon$). We choose $\epsilon$ such that $\textit{K}(L+\epsilon)=1.25\textit{K$_0$}$; this corresponds to about 5\% of the thickness $\textit{d}$. The motivation for this choice is that solutions obtained by setting dark matter sources to zero outside $L+\epsilon$ and using the extended formalism with the diffusion zone boundary at $z=L$ approximately match the solutions obtained from the conventional formalism with the diffusion zone boundary at $z=L+\epsilon$, which suggests that the enhancements must then come from contributions from dark matter sources in the free propagation zone. 
 
 {\it Consistency check ---}
 As an overall consistency check, we also reconstructed the positron density (Eq.~\leqn{Bessel}) in our computation and verified that it satisfies the diffusion-loss equation~\leqn{transport} throughout the diffusion zone for $b=0.$ This was done for all results discussed in this paper.


\begin{thebibliography}{99}

\bibitem{HEAT}
  S.~W.~Barwick {\it et al.}  [HEAT Collaboration],
  Astrophys.\ J.\  {\bf 482}, L191 (1997)
  [arXiv:astro-ph/9703192].

\bibitem{Pamela}
  O.~Adriani {\it et al.}  [PAMELA Collaboration],
  Nature {\bf 458}, 607 (2009)
  [arXiv:0810.4995 [astro-ph]].

\bibitem{ATIC}
  J.~Chang {\it et al.},
  Nature {\bf 456}, 362 (2008).

\bibitem{FERMI}
  A.~A.~Abdo {\it et al.}  [The Fermi LAT Collaboration],
  Phys.\ Rev.\ Lett.\  {\bf 102}, 181101 (2009)
  [arXiv:0905.0025 [astro-ph.HE]].

\bibitem{HESS}
  F.~Aharonian {\it et al.}  [H.E.S.S. Collaboration],
  Phys.\ Rev.\ Lett.\  {\bf 101}, 261104 (2008)
  [arXiv:0811.3894 [astro-ph]];
  H.~E.~S.~Aharonian,
  Astron.\ Astrophys.\  {\bf 508}, 561 (2009)
  [arXiv:0905.0105 [astro-ph.HE]].

\bibitem{Sommer}
  M.~Cirelli, M.~Kadastik, M.~Raidal and A.~Strumia,
  Nucl.\ Phys.\  B {\bf 813}, 1 (2009)
  [arXiv:0809.2409 [hep-ph]];
  N.~Arkani-Hamed, D.~P.~Finkbeiner, T.~R.~Slatyer and N.~Weiner,
  Phys.\ Rev.\  D {\bf 79}, 015014 (2009)
  [arXiv:0810.0713 [hep-ph]].

\bibitem{DecayDM}
  C.~R.~Chen, F.~Takahashi and T.~T.~Yanagida,
  Phys.\ Lett.\  B {\bf 671}, 71 (2009)
  [arXiv:0809.0792 [hep-ph]];
    E.~Nardi, F.~Sannino and A.~Strumia,
  JCAP {\bf 0901}, 043 (2009)
  [arXiv:0811.4153 [hep-ph]];
    A.~Arvanitaki, S.~Dimopoulos, S.~Dubovsky, P.~W.~Graham, R.~Harnik and S.~Rajendran,
  Phys.\ Rev.\  D {\bf 80}, 055011 (2009)
  [arXiv:0904.2789 [hep-ph]].
  
\bibitem{pulsars}
  D.~Hooper, P.~Blasi and P.~D.~Serpico,
  JCAP {\bf 0901}, 025 (2009)
  [arXiv:0810.1527 [astro-ph]];
   H.~Yuksel, M.~D.~Kistler and T.~Stanev,
  Phys.\ Rev.\ Lett.\  {\bf 103}, 051101 (2009)
  [arXiv:0810.2784 [astro-ph]];
   S.~Profumo,
  arXiv:0812.4457 [astro-ph].

\bibitem{Vahe}
  L.~Stawarz, V.~Petrosian and R.~D.~Blandford,
  Astrophys.\ J.\  {\bf 710}, 236 (2010)
  [arXiv:0908.1094 [astro-ph.GA]].

\bibitem{GALPROP}
  I.~V.~Moskalenko and A.~W.~Strong,
  Astrophys.\ J.\  {\bf 493}, 694 (1998)
  [arXiv:astro-ph/9710124];
  A.~W.~Strong, I.~V.~Moskalenko, T.~A.~Porter, G.~Johannesson, E.~Orlando and S.~W.~Digel,
  arXiv:0907.0559 [astro-ph.HE].

\bibitem{Green}
  E.~A.~Baltz and J.~Edsjo,
  Phys.\ Rev.\  D {\bf 59}, 023511 (1998)
  [arXiv:astro-ph/9808243].
  
\bibitem{Bessel}
  T.~Delahaye, R.~Lineros, F.~Donato, N.~Fornengo and P.~Salati,
  Phys.\ Rev.\  D {\bf 77}, 063527 (2008)
  [arXiv:0712.2312 [astro-ph]].

\bibitem{strumia}
  M.~Papucci and A.~Strumia,
  JCAP {\bf 1003}, 014 (2010)
  [arXiv:0912.0742 [hep-ph]].

\bibitem{Grasso}
  C.~Evoli, D.~Gaggero, D.~Grasso and L.~Maccione,
  JCAP {\bf 0810}, 018 (2008)
  [arXiv:0807.4730 [astro-ph]].

\bibitem{Cline}
ÊÊJ.~M.~Cline, A.~C.~Vincent and W.~Xue,
ÊÊ
ÊÊarXiv:1001.5399 [astro-ph.CO].
ÊÊ

\bibitem{CRspectra}
  I.~Cholis, G.~Dobler, D.~P.~Finkbeiner, L.~Goodenough and N.~Weiner,
  Phys.\ Rev.\  D {\bf 80}, 123518 (2009)
  [arXiv:0811.3641 [astro-ph]];
  E.~Borriello, A.~Cuoco and G.~Miele,
  Astrophys.\ J.\  {\bf 699}, L59 (2009)
  [arXiv:0903.1852 [astro-ph.GA]].

\bibitem{BC}
  D.~Maurin, F.~Donato, R.~Taillet and P.~Salati,
  Astrophys.\ J.\  {\bf 555}, 585 (2001)
  [arXiv:astro-ph/0101231].

\bibitem{Patrick}
  P.~Meade, M.~Papucci, A.~Strumia and T.~Volansky,
  Nucl. Phys. B {\bf831}, 178 (2010)  arXiv:0905.0480 [hep-ph].
  
  \bibitem{ICS}
  M.~Cirelli and P.~Panci,
  Nucl.\ Phys.\  B {\bf 821}, 399 (2009)
  [arXiv:0904.3830 [astro-ph.CO]].
  
\bibitem{Portal}
  Y.~Nomura and J.~Thaler,
  Phys.\ Rev.\  D {\bf 79}, 075008 (2009)
  [arXiv:0810.5397 [hep-ph]];
  J.~Mardon, Y.~Nomura, D.~Stolarski and J.~Thaler,
  JCAP {\bf 0905}, 016 (2009)
  [arXiv:0901.2926 [hep-ph]].

\end{thebibliography}
\end{document}